\documentclass[prx,superscriptaddress,aps,amsmath,amssymb,floatfix,reprint,raggedbottom,longbibliography]{revtex4-2}
\usepackage{graphicx}
\usepackage{import}
\usepackage{balance}
\usepackage{xr-hyper}
\usepackage[usenames]{color} 
\usepackage{amssymb} 
\usepackage{caption}
\usepackage[utf8]{inputenc}
\usepackage[table]{xcolor}
\usepackage{colortbl}
\usepackage{float}
\usepackage{braket}
\usepackage{array}
\usepackage{multirow}
\usepackage{booktabs}
\usepackage{amsmath}
\usepackage{ragged2e}
\usepackage{alphalph}

\usepackage{times}

\usepackage{dcolumn}
\usepackage{xcolor}
\usepackage{lipsum}
\usepackage{soul}
\usepackage{braket}
\usepackage{amsfonts}
\usepackage{multirow}
\usepackage[utf8]{inputenc}
\DeclareUnicodeCharacter{2009}{\,}

\usepackage{libertine} 
\usepackage{siunitx}
\usepackage[ruled]{algorithm2e}
\makeatletter
\newcommand*{\addFileDependency}[1]{% argument=file name and extension
  \typeout{(#1)}
  \@addtofilelist{#1}
  \IfFileExists{#1}{}{\typeout{No file #1.}}
}
\makeatother

\makeatletter 
\renewcommand{\fnum@figure}{\textbf{Fig.~\thefigure}}
\makeatother

\newcommand{\beginsupplement}{%
        \setcounter{section}{0}
        \renewcommand{\thesection}{\Roman{section}}
        \setcounter{table}{0}
        \renewcommand{\thetable}{Supplementary Table \arabic{table}}%
        \setcounter{figure}{0}
        \renewcommand{\thefigure}{Supplementary Figure \arabic{figure}}%
        \setcounter{equation}{0}
        \renewcommand{\theequation}{\arabic{equation}}%
     }

\makeatletter
\def\bbordermatrix#1{\begingroup \m@th
  \@tempdima 4.75\p@
  \setbox\z@\vbox{%
    \def\cr{\crcr\noalign{\kern2\p@\global\let\cr\endline}}%
    \ialign{$##$\hfil\kern2\p@\kern\@tempdima&\thinspace\hfil$##$\hfil
      &&\quad\hfil$##$\hfil\crcr
      \omit\strut\hfil\crcr\noalign{\kern-\baselineskip}%
      #1\crcr\omit\strut\cr}}%
  \setbox\tw@\vbox{\unvcopy\z@\global\setbox\@ne\lastbox}%
  \setbox\tw@\hbox{\unhbox\@ne\unskip\global\setbox\@ne\lastbox}%
  \setbox\tw@\hbox{$\kern\wd\@ne\kern-\@tempdima\left[\kern-\wd\@ne
    \global\setbox\@ne\vbox{\box\@ne\kern2\p@}%
    \vcenter{\kern-\ht\@ne\unvbox\z@\kern-\baselineskip}\,\right]$}%
  \null\;\vbox{\kern\ht\@ne\box\tw@}\endgroup}
\makeatother

\emergencystretch=\maxdimen
\usepackage[compact]{titlesec}
\titlespacing{\section}{0pt}{*3}{*2}
\titlespacing{\subsection}{0pt}{*2}{*2}
\titlespacing{\subsubsection}{0pt}{*2}{*2}

\begin{document}

\title{CMOS + stochastic nanomagnets: heterogeneous computers for probabilistic inference and learning}
\par

\affiliation{Department of Electrical and Computer Engineering, University of California Santa Barbara, Santa Barbara, 93106, CA, USA}
\affiliation{Research Institute of Electrical Communication, Tohoku University, 2-1-1 Katahira, Aoba-ku, Sendai 980-8577, Japan}
\affiliation{Graduate School of Engineering, Tohoku University, 6-6 Aramaki Aza Aoba, Aoba-ku, Sendai 980-0845, Japan}

\affiliation{WPI Advanced Institute for Materials Research (WPI-AIMR), Tohoku University, 2-1-1 Katahira, Aoba-ku, Sendai 980-8577, Japan}
\affiliation{Center for Science and Innovation in Spintronics (CSIS), Tohoku University, 2-1-1 Katahira, Aoba-ku, Sendai 980-8577, Japan}
\affiliation{PRESTO, Japan Science and Technology Agency (JST), Kawaguchi 332-0012, Japan}
\affiliation{Division for the Establishment of Frontier Sciences of Organization for Advanced Studies at Tohoku University, Tohoku University, Sendai 980-8577, Japan}
\affiliation{Center for Innovative Integrated Electronic Systems (CIES), Tohoku University, 468-1 Aramaki Aza Aoba, Aoba-ku, Sendai 980-0845, Japan}

\affiliation{Graduate School of Engineering, Tohoku University, 6-6 Aramaki Aza Aoba, Aoba-ku, Sendai 980-0845, Japan}
\affiliation{Research Institute of Electrical Communication, Tohoku University, 2-1-1 Katahira, Aoba-ku, Sendai 980-8577, Japan}
\affiliation{National Institutes for Quantum Science and Technology, Takasaki 370-1207, Japan}
\affiliation{Inamori Research Institute of Science (InaRIS), Kyoto 600-8411, Japan}

\author{Nihal Sanjay Singh}\thanks{These authors contributed equally}
\affiliation{Department of Electrical and Computer Engineering, University of California Santa Barbara, Santa Barbara, 93106, CA, USA}
\author{Keito Kobayashi}\thanks{These authors contributed equally}
\affiliation{Department of Electrical and Computer Engineering, University of California Santa Barbara, Santa Barbara, 93106, CA, USA}
\affiliation{Research Institute of Electrical Communication, Tohoku University, 2-1-1 Katahira, Aoba-ku, Sendai 980-8577, Japan}
\affiliation{Graduate School of Engineering, Tohoku University, 6-6 Aramaki Aza Aoba, Aoba-ku, Sendai 980-0845, Japan}
\author{Qixuan Cao}\thanks{These authors contributed equally}
\affiliation{Department of Electrical and Computer Engineering, University of California Santa Barbara, Santa Barbara, 93106, CA, USA}
\author{Kemal Selcuk}
\affiliation{Department of Electrical and Computer Engineering, University of California Santa Barbara, Santa Barbara, 93106, CA, USA}
\author{Tianrui Hu}
\affiliation{Department of Electrical and Computer Engineering, University of California Santa Barbara, Santa Barbara, 93106, CA, USA}
\author{Shaila Niazi}
\affiliation{Department of Electrical and Computer Engineering, University of California Santa Barbara, Santa Barbara, 93106, CA, USA}
\author{Navid Anjum Aadit}
\affiliation{Department of Electrical and Computer Engineering, University of California Santa Barbara, Santa Barbara, 93106, CA, USA}
\author{Shun Kanai}
\affiliation{Research Institute of Electrical Communication, Tohoku University, 2-1-1 Katahira, Aoba-ku, Sendai 980-8577, Japan}
\affiliation{Graduate School of Engineering, Tohoku University, 6-6 Aramaki Aza Aoba, Aoba-ku, Sendai 980-0845, Japan}
\affiliation{WPI Advanced Institute for Materials Research (WPI-AIMR), Tohoku University, 2-1-1 Katahira, Aoba-ku, Sendai 980-8577, Japan}
\affiliation{Center for Science and Innovation in Spintronics (CSIS), Tohoku University, 2-1-1 Katahira, Aoba-ku, Sendai 980-8577, Japan}
\affiliation{PRESTO, Japan Science and Technology Agency (JST), Kawaguchi 332-0012, Japan}
\affiliation{Division for the Establishment of Frontier Sciences of Organization for Advanced Studies at Tohoku University, Tohoku University, Sendai 980-8577, Japan}
\affiliation{National Institutes for Quantum Science and Technology, Takasaki 370-1207, Japan}

\author{Hideo Ohno}
\affiliation{Research Institute of Electrical Communication, Tohoku University, 2-1-1 Katahira, Aoba-ku, Sendai 980-8577, Japan}
\affiliation{WPI Advanced Institute for Materials Research (WPI-AIMR), Tohoku University, 2-1-1 Katahira, Aoba-ku, Sendai 980-8577, Japan}
\affiliation{Center for Science and Innovation in Spintronics (CSIS), Tohoku University, 2-1-1 Katahira, Aoba-ku, Sendai 980-8577, Japan}
\affiliation{Center for Innovative Integrated Electronic Systems (CIES), Tohoku University, 468-1 Aramaki Aza Aoba, Aoba-ku, Sendai 980-0845, Japan}

\author{Shunsuke Fukami}\email{shunsuke.fukami.c8@tohoku.ac.jp}
\affiliation{Research Institute of Electrical Communication, Tohoku University, 2-1-1 Katahira, Aoba-ku, Sendai 980-8577, Japan}
\affiliation{Graduate School of Engineering, Tohoku University, 6-6 Aramaki Aza Aoba, Aoba-ku, Sendai 980-0845, Japan}
\affiliation{WPI Advanced Institute for Materials Research (WPI-AIMR), Tohoku University, 2-1-1 Katahira, Aoba-ku, Sendai 980-8577, Japan}
\affiliation{Center for Science and Innovation in Spintronics (CSIS), Tohoku University, 2-1-1 Katahira, Aoba-ku, Sendai 980-8577, Japan}
\affiliation{Center for Innovative Integrated Electronic Systems (CIES), Tohoku University, 468-1 Aramaki Aza Aoba, Aoba-ku, Sendai 980-0845, Japan}
\affiliation{Inamori Research Institute of Science (InaRIS), Kyoto 600-8411, Japan}
\author{Kerem Y. Camsari}\email{camsari@ece.ucsb.edu}
\affiliation{Department of Electrical and Computer Engineering, University of California Santa Barbara, Santa Barbara, 93106, CA, USA}
%\date{\today}

\begin{abstract}
%TC:ignore
Extending Moore's law by augmenting complementary-metal-oxide semiconductor (CMOS) transistors with emerging nanotechnologies (X) has become increasingly important. One important class of problems involve sampling-based Monte Carlo algorithms used in probabilistic machine learning, optimization, and quantum simulation. Here, we combine stochastic magnetic tunnel junction (sMTJ)-based probabilistic bits (p-bits) with Field Programmable Gate Arrays (FPGA) to create an energy-efficient CMOS + X (X = sMTJ) prototype. This setup shows how asynchronously driven CMOS circuits controlled by sMTJs can perform probabilistic inference and learning by leveraging the algorithmic update-order-invariance of Gibbs sampling. We show how the stochasticity of sMTJs can augment low-quality random number generators (RNG). Detailed transistor-level comparisons reveal that sMTJ-based p-bits can replace up to 10,000 CMOS transistors while dissipating two orders of magnitude less energy. Integrated versions of our approach can advance probabilistic computing involving deep Boltzmann machines and other energy-based learning algorithms  with extremely high throughput and energy efficiency.
%TC:endignore
\end{abstract}


%\date{}
\maketitle

\section*{}
\label{sec:Intro}
With the slowing down of Moore's Law \cite{theis2017end}, there has been a growing interest in domain-specific hardware and architectures to address emerging computational challenges and energy-efficiency, particularly borne out of machine learning and AI applications. One promising approach is the co-integration of traditional complementary metal-oxide semiconductor (CMOS) technology with emerging nanotechnologies (X), resulting in CMOS + X architectures. The primary objective of this approach is to augment existing CMOS technology with novel functionalities, by enabling the development of physics-inspired hardware systems that realize energy-efficiency, massive parallelism, and asynchronous dynamics, and apply them to a wide range of problems across various domains. 
% By blending CMOS with alternative materials and devices, CMOS + X architectures realize physics-inspired properties can enhance traditional CMOS technologies in terms of energy-efficiency and performance. 

Being named one of the top 10 algorithms of the $20^{\text{th}}$ century \cite{dongarra2000guest}, Monte Carlo methods have been one of the most effective approaches in computing to solve computationally hard problems in a wide range of applications, from probabilistic machine learning, optimization to quantum simulation. Probabilistic computing with p-bits \cite{camsari2017stochasticL} has emerged as a powerful platform for executing these Monte Carlo algorithms in massively parallel \cite{sutton2020autonomous,aadit2022massively} and energy-efficient architectures. p-bits have been shown to be applicable to a large domain of computational problems from  combinatorial optimization to probabilistic machine learning and quantum simulation \cite{kaiser2021probabilistic,camsari2019p,
chowdhury2023full}. 

\begin{figure*}[t!]
    \centering    \includegraphics[width=0.825\textwidth]{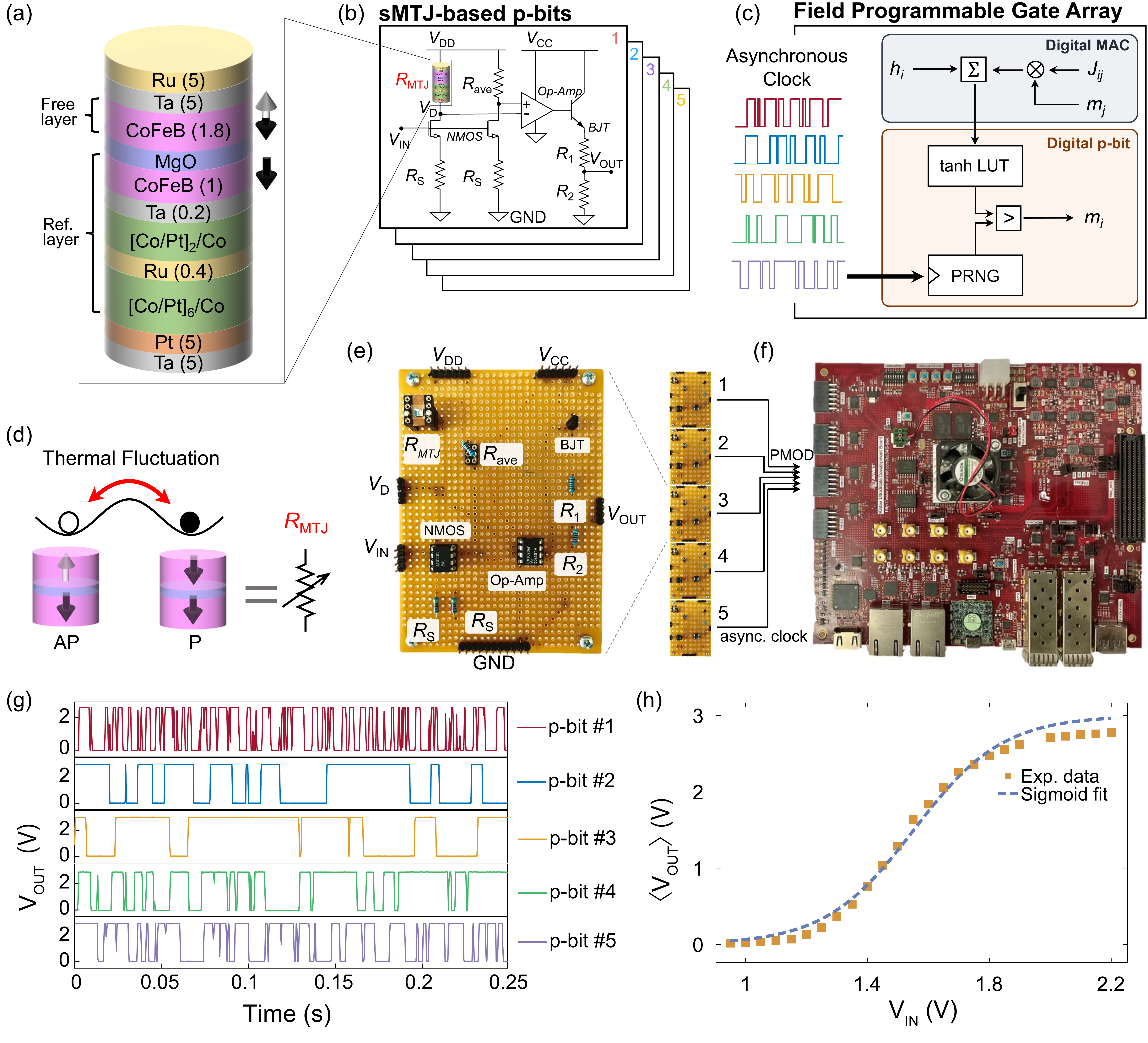}
    \caption{\textbf{Experimental setup for the CMOS + sMTJ  probabilistic computer.} (a) Stack structure of the stochastic magnetic tunnel junction (sMTJ). (b) The proposed sMTJ-based p-bit circuit with two branches whose outputs are provided to an operational amplifier. $R_{\text{ave}}$ is the average resistance of ${R_\text{P}}$ and $R_{\text{AP}}$ of the sMTJ.  5 sMTJ-based p-bits provide tunable, truly random and asynchronous clocks to a digital field programmable gate array (FPGA). (c) Digital p-bits in the FPGA use lookup tables (LUT), comparators, synaptic weights, and pseudorandom number generators (PRNG). The clocks of the PRNG are driven by the truly random asynchronous outputs coming from the analog p-bits. (d) Pictorial representation of perpendicular sMTJ. (e) Image of a single p-bit circuit. (f) Image of the FPGA. The asynchronous clocks are input through the peripheral module (PMOD) pins. (g) Typical output of p-bits \#1 to 5 using 5 sMTJs obtained from the p-bit circuit (see Supplementary Section~\ref{sec:pcirc}), showing variations in fluctuations. (h) Experimentally measured $\langle V_\text{OUT} \rangle$ the time average (over a period of 3 minutes) of the p-bit circuit output, as a function of DC input voltage $V_\text{IN}$. The yellow squares are experimental data, and the blue dashed line is a fit of the form $\langle V_\text{OUT} \rangle = 1/2 \ V'_\text{CC} [\tanh[\beta(V_\text{IN} - V_0)] + 1]$, where $V_0 =1.55 \ \si{V}$, $\beta=3.43 \ \si{V}^{-1}$,  $V'_\text{CC} =3\ \si{V}$ is a reduced voltage from $V_{\text{CC}}=5\ \si{V}$ (see Supplementary Section~\ref{sec:expcirc}).}
     \label{fig:Fig1}
\end{figure*}

Several p-bit implementations that use the inherent stochasticity in different materials and devices have been proposed, based on diffusive memristors \cite{woo2022probabilistic}, resistive RAM \cite{liu2022probabilistic}, perovskite nickelates \cite{park2022efficient}, ferroelectric transistors \cite{luo2023probabilistic}, single photon avalanche diodes \cite{whitehead2023cmos}, optical parametric oscillators \cite{roques2023biasing} and others. Among alternatives sMTJs built out of low-barrier nanomagnets have demonstrated significant potential due to their ability to amplify noise, converting millivolts of fluctuations to hundreds of millivolts over resistive networks \cite{camsari2017implementing}, unlike alternative approaches with amplifiers \cite{cheemalavagu2005probabilistic}. Another advantage of sMTJ-based p-bits is the continuous generation of truly random bitstreams without the need to be reset in synchronous pulse-based designs \cite{fukushima2014spin,Rehm2023transducer}. The possibility of designing energy-efficient p-bits using low-barrier nanomagnets has stimulated renewed interest in material and device research with several exciting demonstrations from nanosecond fluctuations  \cite{safranski2021demonstration,hayakawa2021nanosecond,schnitzspan2023nanosecond} to better theoretical understanding of nanomagnet physics \cite{kaiser2019subnanosecond,hassan2019low,kanai2021theory,funatsu2022local} and novel magnetic tunnel junction designs \cite{camsari2021double,kobayashi2022external}. 

Despite promising progress with hardware  prototypes \cite{borders2019integer,kaiser2022hardware,si2023energy,gibeault2023programmable,daniel2023experimental}, large-scale probabilistic computing using stochastic nanodevices remains elusive.  As we will establish in this paper, designing purely CMOS-based high-performance probabilistic computers suited to sampling and optimization problems is prohibitive beyond a certain scale ($>$1M p-bits) due to the large area and energy costs of pseudorandom number generators. As such, any large-scale integration of probabilistic computing will involve strong integration with CMOS technology in the form of CMOS+X architectures. Given the unavoidable device-to-device variability, the interplay between  continuously fluctuating stochastic nanodevices (e.g., sMTJs) with deterministic CMOS circuits and possible applications of such hybrid circuits remain unclear. 

In this paper, we {first} introduce the notion of a heterogeneous CMOS+sMTJ system where the asynchronous dynamics of sMTJs control digital circuits in a standard CMOS Field Programmable Gate Array (FPGA). We view the FPGA as a ``drop-in replacement'' for eventual integrated circuits where sMTJs could be situated on top of CMOS. Unlike earlier implementations where sMTJs were primarily used to implement neurons and CMOS or analog components circuits for synapses \cite{borders2019integer,kaiser2022hardware}, we design hybrid circuits where sMTJ-based p-bits control a large number of digital circuits residing in the FPGA without dividing the system into neurons (sMTJ) and synapses (CMOS). We show how the true randomness injected into deterministic CMOS circuits augment low-quality random number generators based on linear feedback shift registers (LFSR). This result represents an example of how sMTJs could be used to reduce footprint and energy consumption in the CMOS underlayer. In this work, we present a  small example of a CMOS + sMTJ system, however, similar systems can be scaled up to much bigger densities, leveraging the proven manufacturability of magnetic memory at gigabit densities. Our results will help lay the groundwork for larger implementations in the presence of unavoidable device-to-device variations. We also focus beyond the common use case of combinatorial optimization of similar physical computers \cite{mohseni2022ising}, considering probabilistic inference and learning in deep energy-based models.  

Specifically, we use our system to train 3-hidden 1-visible layer deep and unrestricted Boltzmann machines that entirely rely on the asynchronous dynamics of the stochastic MTJs. Second, we evaluate the quality of randomness directly at the application level through probabilistic inference and deep Boltzmann learning.  This approach contrasts with the majority of related work, which typically conducts statistical tests at the single device level to evaluate the quality of randomness \cite{vodenicarevic2017low,parks2018superparamagnetic,schnitzspan2023nanosecond,ostwal2019spin,lv2022bipolar,fu2021overview} (see Supplementary Sections~\ref{sec:randperm_supp}, \ref{sec:bias}, and \ref{sec:NIST} for more randomness experiments). As an important new result, we find that the quality of randomness matters in machine learning tasks as opposed to optimization tasks that have been explored previously. And finally, we conduct a comprehensive benchmark using an experimentally calibrated 7-nm CMOS PDK and find that when the quality of randomness is accounted for, the sMTJ-based p-bits are about 4 orders of magnitude smaller in area and they dissipate 2 orders of magnitude less energy, compared to CMOS p-bits. We envision that large-scale CMOS+X p-computers ($\gg 10^5$) can be a reality in scaled up versions of the CMOS + sMTJ type computers we discuss in this work. 

\begin{figure*}[t!]
    \centering
    \includegraphics[width=0.82\linewidth]{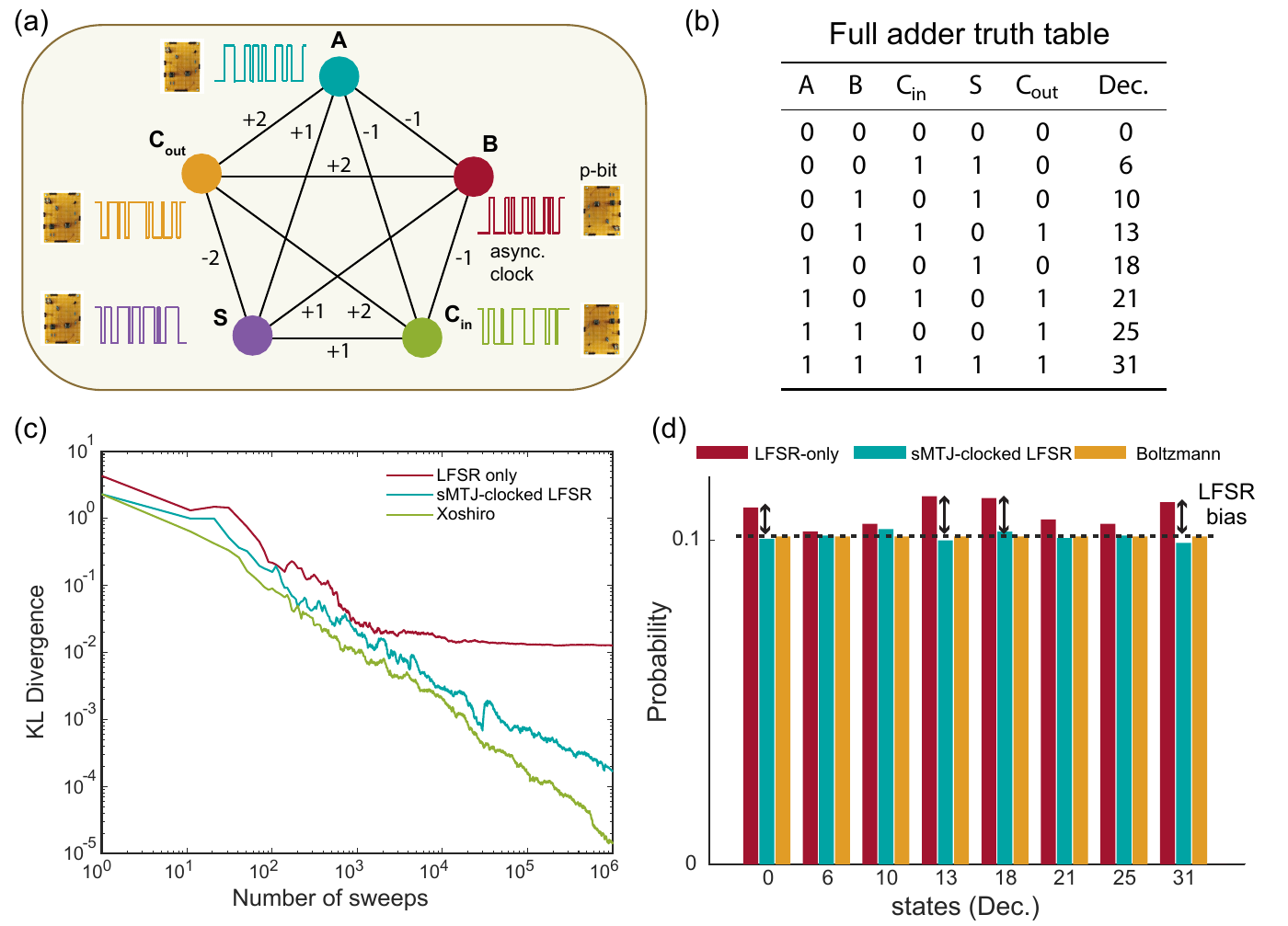}
    \caption{\textbf{Inference on a probabilistic full adder.} (a) Fully-connected full adder network \cite{smithson2019efficient}, where p-bits are clocked by the sMTJs. (b) Truth table of the full adder where Dec. represents the decimal representation of the state of [$\rm A$ $\rm B$ $\rm C_{\text{in}}$ $\rm S$ $\rm C_{\text{out}}$] from left to right. (c) Kullback-Leibler (KL) divergence between the ideal and measured distributions vs. the number of sweeps. Results are shown for LFSR-based p-bit (red line), sMTJ-clocked LFSR-based p-bit (blue line), and Xoshiro-based p-bit (green line). (d) Histogram for the measured and ideal distributions at  the $10^{6}$ sweep. The red, blue, and yellow bars show LFSR, sMTJ-clocked LFSR, and Boltzmann distribution, respectively. The histogram shows all 8 high probability states denoted in (b) and with a clear bias for the LFSR distribution (see Supplementary Section~\ref{sec:learningsupp} for full histograms for all PRNGs including Xoshiro).}
    \label{fig:Fig2}
 \end{figure*}

\section*{Constructing the heterogeneous p-computer}
\label{sec:Experimental setup}

FIG.~\ref{fig:Fig1} shows a broad overview of our sMTJ-FPGA setup along with device and circuit characterization of sMTJ p-bits. Unlike earlier p-bit demonstrations with sMTJs as standalone stochastic binary neurons, in this work, we use sMTJ-based p-bits to generate asynchronous and truly random clock sources to drive digital p-bits in the FPGA (FIG.~\ref{fig:Fig1}a,b,c).

  The conductance of the sMTJ  depends on the relative angle $\theta$ between the free and the fixed layers, $G_{\text{MTJ}}~$~$\propto$~$[1~+~P^2 ~\cos(\theta)]$, where $P$ is the interfacial spin polarization. When the free layer is made out of a low barrier nanomagnet $\theta$ becomes a random variable in the presence of thermal noise, causing conductance fluctuations between the parallel (P) and the antiparallel (AP) states (FIG. \ref{fig:Fig1}d). 
 
 The five sMTJs used in the experiment are designed with a diameter of 50 nm and have a relaxation time of about $1$ to $20$ ms, with  energy barriers of $\approx$14-17 $k_BT$, assuming an attempt time of 1 ns \cite{coffey2012thermal} (see Supplementary Section~\ref{sec:MTJ}). In order to convert these conductance fluctuations into voltages, we design a new p-bit circuit (FIG.~\ref{fig:Fig1}b,e). This circuit creates a voltage comparison between two branches controlled by two transistors, fed to an operational amplifier. As we discuss in Supplementary Section~\ref{sec:pcirc}, the main difference of this circuit compared to the earlier 3 transistor/1MTJ design used in earlier demonstrations \cite{borders2019integer,kaiser2022hardware} is in its ability to provide a larger stochastic window to tune the p-bit (FIG.~\ref{fig:Fig1}h) with more variation tolerance (see Supplementary Section~\ref{sec:expcirc}).

FIG.~\ref{fig:Fig1}f,g show how the asynchronous clocks obtained from p-bits with 50/50 fluctuations are fed to the FPGA. Inside the FPGA, we design a digital probabilistic computer where a p-bit includes a lookup table (LUT) for the hyperbolic tangent function, a pseudorandom number generator (PRNG) and a digital comparator (see Supplementary Section~\ref{sec:FPGA}). 

The crucial link between analog p-bits and the digital FPGA is established through the clock of the PRNG used in the FPGA, where a multitude of digital p-bits can be asynchronously driven by analog p-bits. As we discuss in Sections~3-4, depending on the quality of the chosen PRNG, the injection of additional entropy through the clocks has a considerable impact on inference and learning tasks. The potential for enhancing low-quality PRNGs using compact and scalable nanotechnologies, such as sMTJs, which can be integrated as a BEOL (Back-End-Of-Line) process on top of the CMOS logic, holds significant promise for future CMOS + sMTJ  architectures.

\section*{RESULTS}

\section*{Probabilistic inference with heterogeneous p-computers}
\label{sec:Probabilistic Sampling}

In the p-bit formulation, we define  probabilistic inference as generating samples from a specified distribution which is the  Gibbs-Boltzmann distribution for a given network (see Supplementary Section~\ref{sec:basic} for details). This is a computationally hard problem \cite{goodfellow2016deep}, and is at the heart of many important applications involving  Bayesian inference \cite{friedman2003being}, training probabilistic models in machine learning \cite{long2010restricted}, statistical physics \cite{krauth2006statistical} and many others \cite{andrieu2003introduction}. Due to the broad applicability of probabilistic inference, improving key figures of merit such as probabilistic flips per second (sampling throughput) and energy-delay product for this task are extremely important.

To demonstrate this idea, we evaluate probabilistic inference on a probabilistic version of the full adder (FA) \cite{smithson2019efficient} as shown in FIG.~\ref{fig:Fig2}a. The truth table of the FA is given in FIG.~\ref{fig:Fig2}b. The FA performs 1-bit binary addition and it has three inputs (A, B, Carry in=C$_{\text{in}}$) and two outputs (Sum=S,  and Carry out=C$_{\text{out}}$). The probabilistic FA can be described in a 5 p-bit, fully-connected network (FIG.~\ref{fig:Fig2}a). When the network samples from its equilibrium, it samples states corresponding to the truth table, according to the Boltzmann distribution. 

We demonstrate probabilistic sampling on the probabilistic FA using the digital p-bits with standalone  Linear Feedback Shift Registers LFSRs (only using the FPGA), sMTJ-clocked LFSRs (using sMTJ-based p-bits and the FPGA), and standalone Xoshiro RNGs (only using the FPGA). Our main goal is to compare the quality of randomness obtained by inexpensive but low-quality PRNGs such as LFSRs \cite{paar2009understanding} with sMTJ-enhanced LFSRs and high-quality but expensive PRNGs such as Xoshiro \cite{blackman2021scrambled} (see Supplementary Section~\ref{sec:samplingsupp}).

FIG.~\ref{fig:Fig2}c shows the comparison of these three different solvers where we measure the Kullback-Leibler (KL) divergence \cite{10.1214/aoms/1177729694} between the cumulative distribution based on the number of sweeps and the ideal Boltzmann distribution of the FA: \begin{equation}
 \label{eq:KL}
  \mathrm{KL}[P_\text{exp} \mid\mid P_\text{ideal}]= \sum_{x}{P_\text{exp}(x) \text{log}\frac{
  P_\text{exp}(x)}{P_\text{ideal}(x)}}, 
 \end{equation}
where $P_\text{exp}$ is the probability obtained from the experiment (cumulatively measured) and $P_\text{ideal}$ is the probability obtained from the Boltzmann distribution. For LFSR (red line), the KL divergence saturates when the number of sweeps exceeds $N = 10^4$, while for sMTJ-clocked LFSR (blue line) and Xoshiro (green line), the KL divergence decreases with increasing the number of sweeps. The persistent bias of the LFSR is also visible in the partial histogram of probabilities measured at $N=10^6$ sweeps as shown in FIG.~\ref{fig:Fig2}d (see Supplementary Section~\ref{sec:learningsupp} for the full histograms). It is important to note here in our present context where sMTJs are limited to a handful of devices, we use sMTJ-based p-bits used to drive low-quality LFSRs, observing how they perform similar to high-quality PRNGs. In integrated implementations however, sMTJ-based p-bits can be directly used as p-bits themselves, without any supporting PRNG (see Supplementary Section~\ref{sec:roadmap} for details on projections of integrated implementations).

The mechanism of how the sMTJ-clocked LFSRs produce random numbers is interesting: even though the next bit in an LFSR is always perfectly determined, the randomness in the arrival times of clocks from the sMTJs makes their output unpredictable. Over the course of the full network's evolution, each LFSR produces an unpredictable bitstream, functioning as truly random bits.

\begin{figure*}[t!]
\includegraphics[width=0.825\textwidth]{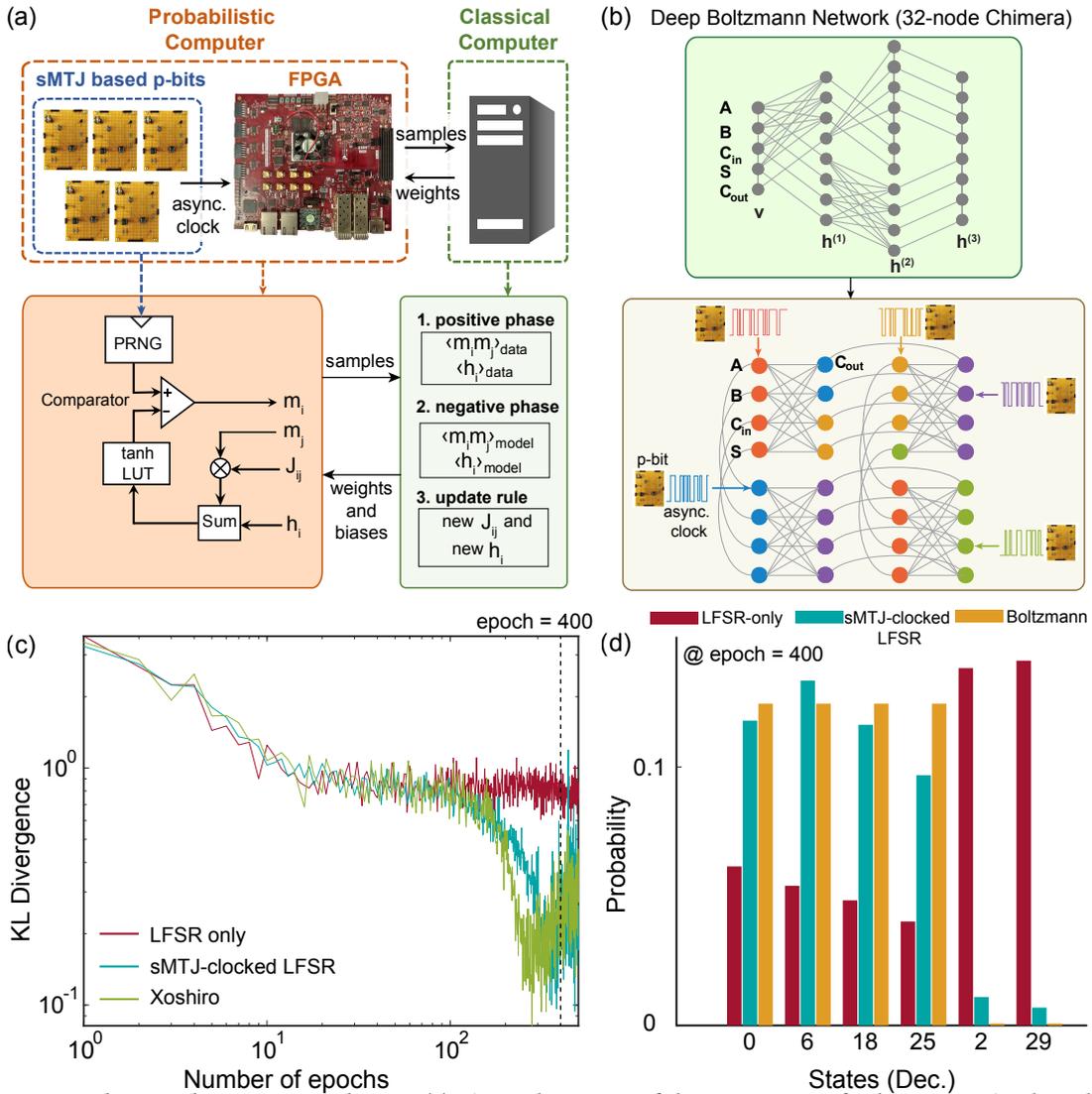}
\vspace{-12.5pt} 
\caption{\textbf{Learning deep Boltzmann machines.}
    (a) The architecture of the p-computer for learning. The digital p-bits in FPGA are fed by sMTJ-based p-bits output similar to probabilistic inference. The weights $J_{ij}$ and biases $h_i$ are updated in the CPU for a specified number of epochs. (b) (Top) The 32-node Chimera graph is used as a deep BM. (Bottom)  An asynchronous clocking scheme is shown with node coloring. (c) KL divergence as a function of the number of epochs for LFSR (red line), LFSR clocked by sMTJ-based p-bit (blue line), and Xoshiro (green line) (d) The distribution of full adder with learned weights and biases at epoch = 400 where the number of sweeps per epoch = 400 for LFSR-only and the number of sweeps per epoch = 16000 for sMTJ-clocked LFSR. The Boltzmann distribution was obtained with $\beta=3$. The red, blue, and yellow bars show LFSR and LFSR clocked by sMTJ-based p-bit, and Boltzmann, respectively. The histogram shows 4 correct (0, 6, 18, 25) and 2 incorrect (2, 29) states, out of the 32 possible states. sMTJ-based p-bit closely approximates the ideal Boltzmann distribution whereas the LFSR underestimates correct states and completely fails with states 2 and 29 (see Supplementary Section~\ref{sec:learningsupp} for full histograms for all PRNGs including Xoshiro).}
\label{fig:Fig3}
\vspace{-12.5pt} 
 \end{figure*}

The observed bias of the LFSR can be due to several reasons: first, the LFSRs generally provide low-quality random numbers and do not pass all the tests in the NIST statistical test suite \cite{Rukhin2010nist} (see Supplementary Section~\ref{sec:NIST}). Second, we take whole words of random bits from the LFSR to generate large random integers. This is a known danger when using LFSRs \cite{press1988numerical,knuth1981art}, which can be mitigated by the use of phase shifters that scramble the parallelly obtained bits to reduce their correlation \cite{rajski1998design}. But such measures increase the complexity of PRNG designs further limiting the scalable implementation of digital p-computers (see Supplementary Section~\ref{sec:bias} for detailed experimental analysis of LFSR bias). 

The quality of randomness in Monte Carlo sampling is a rich and well-studied subject (see, for example, \cite{PARISI1985301,FILK1985125,PhysRevLett.73.2513}). The main point we stress in this work is that even compact and inexpensive simple PRNGs can perform as well as sophisticated, high-quality RNGs when augmented by truly random nanodevices such as sMTJs. 
%abc:endignore

\section*{Boltzmann Learning with heterogeneous p-computers}
\label{sec:Machine Learning}

We now show how to train deep Boltzmann machines (DBM) with our heterogeneous sMTJ + FPGA-based computer. Unlike probabilistic inference, in this setting, the weights of the network are unknown and the purpose of the training process is to obtain desired weights for a given truth table, such as the full adder (see Supplementary Section~\ref{sec:arbitrary} for an example of arbitrary distribution generation using the same learning algorithm).  We consider this demonstration as a proof-of-concept  for eventual larger-scale implementations (FIG.~\ref{fig:Fig3}a,b). Similar to probabilistic inference, we compare the performance of three solvers: LFSR-based, Xoshiro-based and sMTJ+LFSR-based RNGs. We choose a 32-node Chimera lattice \cite{boothby2020next} to train a probabilistic full adder with 5 visible nodes and 27 hidden nodes in a 3-layer DBM (see FIG.~\ref{fig:Fig3}b top panel). Note that this deep network is significantly harder to train than training fully-visible networks whose data correlations are known a priori \cite{kaiser2022hardware}, necessitating positive and negative phase computations (see Supplementary Section~\ref{sec:learningsupp} and Algorithm~\ref{alg:alg2} for details on the learning algorithm and implementation).

FIG.~\ref{fig:Fig3}c,d show the KL divergence and the probability distribution of the full adder Boltzmann machines based on the fully digital LFSR/Xoshiro and the heterogeneous sMTJ-clocked LFSR RNGs. The KL divergence in the learning experiment is performed like this: after each epoch during training, we save the weights in the classical computer and perform probabilistic inference to measure the KL distance between the learned and ideal distributions.  The sMTJ-clocked LFSR and the Xoshiro based Boltzmann machines produce probability distributions that eventually closely approximate the Boltzmann distribution of the full adder. On the other hand, the fully digital LFSR based Boltzmann machine produces the incorrect states $\rm [A~B~C_{\text{in}}~S~C_{\text{out}}] = 2$ and $29$ with a significantly higher probability than the correct peaks, and grossly underestimates the probabilities of states 0, 6, 18, and 25 (see FIG.~\ref{fig:FigS4} for full histograms that are avoided here for clarity). As in the inference experiment  (FIG.~\ref{fig:Fig2}a), the KL divergence of the LFSR saturates and never improves beyond a point. The increase in the KL divergence for Xoshiro and sMTJ-clocked LFSR towards the end is related to hyperparameter selection and unrelated to RNG quality \cite{dabelow2022three}. For this reason, we select the weights at epoch=400 for testing to produce the histogram in FIG.~\ref{fig:Fig3}d. 

In line with our previous results, the learning experiments confirm the inferior quality of LFSR-based PRNGs, particularly for learning tasks (see Supplementary Section~\ref{sec:trainDBM_supp} for an MNIST training comparisons between p-bits based on Xoshiro and LFSR). While LFSRs can produce correct peaks with some bias in optimization problems, they fail to learn appropriate weights for sampling and learning, rendering them unsuitable for these applications. In addition to these results, statistical tests on the NIST test suite corroborate our findings that sMTJ-clocked LFSRs and high-quality PRNGs such as Xoshiro outperform the pure LFSR-based p-bits (see Supplementary Section~\ref{sec:NIST}).

Our learning result demonstrates how asynchronously interacting p-bits can creatively combine with existing CMOS technology. Scaled and integrated implementations of this concept could lead to a resurgence in training powerful deep Boltzmann machines \cite{niazi2023training}.

\begin{figure*}[t!]
    \centering
    \includegraphics[width=0.85\textwidth]{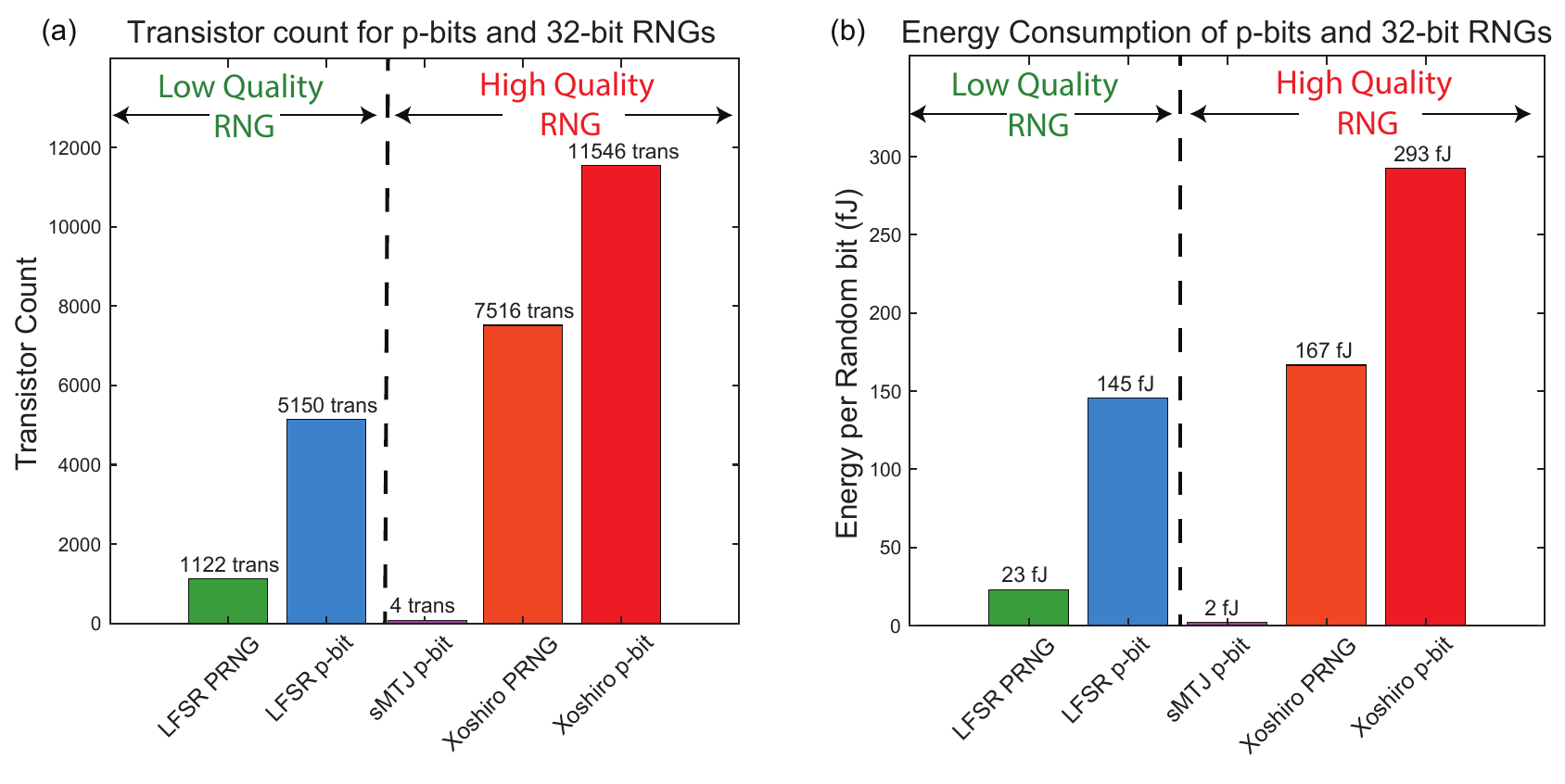}
\vspace{-15pt}\caption{\textbf{Transistor counts and energy consumption for p-bit and RNG implementations} 
    The digital p-bits and PRNGs are synthesized by the ASAP7 PDK and simulated in HSPICE in transistor-level simulations. The PRNGs are 32-bits long and LUTs store $2^8$ words that are 32-bits long to be compared with 32-bit RNGs. The sMTJ-based p-bit result is repeated from \cite{borders2019integer}. To activate the LUT, a periodic input signal with low inputs to the p-bit has been used. See the text and  Supplementary Section~\ref{sec:energy} for details on the energy calculation.}  
\label{fig:Fig4}
 \end{figure*}

\section*{Energy and transistor count comparisons }
\label{sec:benchmark}

Given our prior results stressing how the quality of randomness can play a critical role in probabilistic inference and learning, it is beneficial to perform precise, quantitative comparisons with the various digital PRNGs we built in hardware FPGAs with sMTJ-based p-bits \cite{camsari2017implementing}. Note that for this comparison, we do not consider augmented CMOS p-bits, but directly compare sMTJ-based mixed signal p-bits with their digital counterparts (see Supplementary Section~\ref{sec:roadmap} for details on projections of integrated implementations using sMTJ-based mixed signal p-bits). Moreover, instead of benchmarking the voltage comparator based p-bit circuit shown in FIG.~\ref{fig:Fig1} or other types of spin-orbit torque based p-bits \cite{camsari2017stochasticL,yin2022scalable}, we benchmark the 3T/1MTJ based p-bit first reported in \cite{camsari2017implementing}. The reason for this choice is that this design allows the use of fast in-plane sMTJs whose fluctuations can be as fast as micro to nanoseconds. We also note that the table-top components we use in this work are not optimized but used for convenience.

For the purpose of benchmarking and characterization, we synthesize circuits for LFSR and Xoshiro PRNGs and these PRNG-based p-bits using the ASAP 7nm Predictive process development kit (PDK) that uses SPICE-compatible FinFET device models \cite{CLARK2016105}. Our synthesis flow, explained in detail in Supplementary Section~\ref{sec:synth}, starts from hardware description level (HDL) coding of these PRNGs and leads to transistor-level circuits using the experimentally benchmarked ASAP 7nm PDK. As such, the analysis we perform here offers a high degree of precision in terms of transistor counts and quantitative energy consumption. 

FIG.~\ref{fig:Fig4}a shows the transistor count for p-bits using 32-bit PRNGs. Three pieces make up a digital p-bit: PRNG, LUT (for the activation function) and a digital comparator (typically small). To understand how each piece contributes to the transistor count, we separate the PRNG from the LUT contributions in FIG.~\ref{fig:Fig4}a. 

First, we reproduce earlier results reported in Ref.~\cite{borders2019integer}, corresponding to the benchmarking of the design reported in \cite{camsari2017implementing} and find that a 32-bit LFSR requires 1122 transistors which is very close to the custom-designed 32-bit LFSR with 1194 transistors in Ref. \cite{borders2019integer}. However, we find that the addition of a LUT, ignored in \cite{borders2019integer}, adds significantly more transistors. Even though the inputs to the p-bit are 10-bits (s[6][3]), the saturating behavior of the $\text{tanh}$ activation allows reductions in LUT size.  In our design, the LUT stores $2^8$ words of 32-bit length that are compared to the 32-bit PRNG. Under this precision, the LUT increases the transistor count to 5150, and more would be needed for finer representations. Note that the compact sMTJ-based p-bit design proposed in \cite{camsari2017implementing} uses 3 transistors plus an sMTJ which we estimate as having an area of 4 transistors, following Ref.~\cite{borders2019integer}. In this case, there is no explicit need for a LUT or a PRNG. 

Additionally, the results presented in FIG.~\ref{fig:Fig2} and ~\ref{fig:Fig3}  indicate that to match the performance of the sMTJ-based p-bits, more sophisticated PRNGs like Xoshiro must be used. In this case, merely the PRNG cost of a 32-bit Xoshiro is 7516 transistors. The LUT costs are the same as LFSR-based p-bits which is about $\approx$ 4029 transistors. 

Collectively, these results indicate that to truly replicate the performance of an sMTJ-based p-bit, the actual transistor cost of a digital design is about 11,000 transistors which is an order of magnitude worse than the conservative estimation performed in Ref.~\cite{borders2019integer}. 

In FIG.~\ref{fig:Fig4}b we show the energy costs of these differences. We focus on the energy required to produce one random bit. Once again, our synthesis flow, followed by ASAP7 based HSPICE simulations reproduces the results presented in Ref.~\cite{borders2019integer}. We estimate a 23 fJ energy per random bit from the LFSR based PRNG where this number was reported to be 20 fJ in Ref.~\cite{borders2019integer}.

Similar to the transistor count analysis, we consider the effect of the LUT on the energy consumption which was absent in \cite{borders2019integer}. We first observe that if the LUT is not active, i.e., if the input $I_i$ to the p-bit is not changing, the LUT does not change the energy per random bit very much.  In a real p-circuit computation, however, $I_i$ would be continuously changing activating the LUT repeatedly. To simulate these working conditions, we create a variable $I_i$ pulse that wanders around the stochastic window of the p-bit by changing the least significant bits of the input (see Supplementary Section~\ref{sec:energy}). We choose a 1 GHz frequency for this pulse mimicking an sMTJ with a lifetime of 1 ns. We observe that in this case the total energy to create a random bit on average increases by a factor of 6$\times$ for the LFSR, reaching 145 fJ per bit.

For the practically more relevant Xoshiro, the average consumption per random bit reaches around 293 fJ. Once again, we conclude that the 20 fJ per random bit, reported in Ref.~\cite{borders2019integer} underestimates the costs of RNG generation by about an order of magnitude when the RNG quality and other peripheries such as LUTs are carefully taken into account. In this paper, we do not reproduce the energy estimation of the sMTJ-based p-bit but report the estimate in Ref.~\cite{borders2019integer} which assumes an sMTJ-based p-bit with $\approx$ nanosecond fluctuations. 

Our benchmarking results highlight the true expense of high-quality, digital p-bits in silicon implementations. Given that functionally interesting and sophisticated p-circuits require above 10000 to 50000 p-bits \cite{aadit2022massively}, using a 32-bit Xoshiro-based p-bit in a digital design would consume up to \emph{0.1 to 0.5 Billion transistors}, just for the p-bits. In addition, the limitation of not being able to parallelize or fit more random numbers in hardware would limit the throughput \cite{Misra2022} and the probabilistic flips per second, a key metric measuring the effective sampling speed of a probabilistic computer (see for example, \cite{preis2009gpu,yang2019high,romero2020high}). 
% New Addition
{As discussed in detail in Supplementary Section}~\ref{sec:roadmap}, near-term projections with  $N=10^4$ p-bits using sMTJs with in-plane magnetic anisotropy (IMA) ($\tau\approx1$ns~\cite{safranski2021demonstration}) can reach $\approx10^4$ flips/ns in sampling throughput.
These results clearly indicate that a digital solution beyond 10000 to 50000 p-bits, as required by large-scale optimization,  probabilistic machine learning, and optimization tasks, will remain prohibitive. To solve these traditionally expensive but practically useful problems, the heterogeneous integration of sMTJs holds great promise both in terms of scalability and energy-efficiency.

\section*{DISCUSSIONS}
\label{sec:conclusions}

This work demonstrates the first hardware demonstration of a heterogeneous computer combining versatile FPGAs with stochastic MTJs for probabilistic inference and deep Boltzmann learning. We introduce a new variation tolerant p-bit circuit that is used to create an asynchronous clock domain, driving digital p-bits in the FPGA. In the process, the CMOS + sMTJ computer shows how commonly used and inexpensive PRNGs can be augmented by magnetic nanodevices to perform as well as high-quality PRNGs (without the resource overhead), both in probabilistic inference and learning experiments. Our CMOS + sMTJ computer also shows the first demonstration of training a deep Boltzmann network in a 32-node Chimera topology, leveraging the asynchronous dynamics of sMTJs. Careful comparisons with existing digital circuits show the true potential of integrated sMTJs which can be scaled up to million p-bit densities far beyond the capabilities of present day CMOS technology (see Supplementary Section ~\ref{sec:roadmap} for detailed benchmarking and a p-computing roadmap).

\section*{METHODS}

\subsection*{sMTJ fabrication and circuit parameters}
 We employ a conventional fixed and free layer sMTJ, both having perpendicular magnetic anisotropy. The reference layer thickness is 1 nm (CoFeB) while the free layer is 1.8 nm (CoFeB), deliberately made thicker to reduce its energy barrier \cite{borders2019integer,parks2018superparamagnetic}. The stack structure of the sMTJs we use is, starting from the substrate side, Ta(5)/ Pt(5)/ [Co(0.4)/Pt(0.4)]$_6$/ Co(0.4)/ Ru(0.4)/ [Co(0.4)/Pt(0.4)]$_2$/ Co(0.4)/ Ta(0.2)/ CoFeB(1)/ MgO(1.1)/ CoFeB(1.8)/ Ta(5)/ Ru(5), where the numbers are in nanometers (FIG.~\ref{fig:Fig1}a). Films are deposited at room temperature by dc/rf magnetron sputtering on a thermally oxidized Si substrate. The devices are fabricated into a circular shape with a 40-80 nm diameter using electron beam lithography and Ar ion milling and annealed at 300 \si{\degreeCelsius}  for 1 hour by applying a 0.4 T magnetic field in the perpendicular direction. The average tunnel magnetoresistance ratio (TMR) and resistance area product ($\rm RA$) are 65\% and 4.7 $\si{\Omega\um^2}$, respectively. The discrete sMTJs used in this work are first cut out from the wafer and the electrode pads of the sMTJs are bonded with wires to IC sockets.
The following parameters are measured by sweeping DC current to the sMTJ and measuring the voltage. The resistance of the P state $R_\textrm{P}$ is 4.4-5.7 $\si{\kohm}$, the resistance of the AP state $R_\textrm{AP}$ is 5.9-7.4 \si{\kohm}, and the current at which P/AP fluctuations are 50\%  is defined as $I_{50/50}$, in between 14-20  $\mu$A. At the output of the new p-bit design, we use an extra branch  with a bipolar junction transistor (BJT) that acts as a buffer to the peripheral module (PMOD) pins of the Kintex UltraScale KU040 FPGA board. Given the electrostatic sensitivity of the sMTJs, this branch also protects the circuit from any transients that might originate from the FPGA.

\subsection*{Digital synthesis flow}

HDL codes are converted to gate-level models using the Synopsys Design Compiler. Conversion from these models to Spice netlists is done using Calibre Verilog-to-LVS. Netlist post-processing is done by a custom Mathematica script to make it HSPICE compatible. Details of the synthesis flow (shown in FIG.~\ref{fig:Fig4}), followed by HSPICE simulation results for functional verification and power analysis are provided in Supplementary Sections~\ref{sec:synth}, ~\ref{sec:funcver} and ~\ref{sec:energy}. 

\vspace{-5pt} 
\section*{DATA AVAILABILITY}
All processed data generated in this study are provided in the main text and Supplementary Information.
The data that support the plots within this paper and other findings of this study are available from the corresponding author upon request.
\vspace{-5pt} 

\section*{CODE AVAILABILITY}
The computer code used in this study is available from the corresponding author upon request.

\vspace{-5pt} 
\section*{ACKNOWLEDGEMENTS}
 We are grateful to Subhasish Mitra and Carlo Gilardi for discussions regarding Linear Feedback Shift Registers and high-level synthesis. We gratefully acknowledge Kevin Cao and Mishel Jyothis Paul for their help with the configuration of ASAP7 PDK. We are grateful to Shuvro Chowdhury for his comments on an earlier version of this manuscript. 
 The U.S. National Science Foundation (NSF) grant CCF 2106260, the Office of Naval Research Young Investigator Program (YIP) grant, SAMSUNG Global Research Outreach (GRO) grant, and an NSF CAREER grant are acknowledged by N.S.S., Q.C, K.S., T.H., S.N., N.A.A., and K.Y.C for supporting this research.
 % This work is supported in part by a U.S. National Science Foundation (NSF) grant CCF 2106260, the Office of Naval Research Young Investigator Program (YIP) grant, SAMSUNG Global Research Outreach (GRO) grant, and an NSF CAREER grant. 
Murata Science Foundation and Marubun Research Promotion Foundation are acknowledged by K.K. JST-CREST Grant No. JPMJCR19K3, JST-AdCORP Grant No. JPMJKB2305, and MEXT X-NICS Grant No. JPJ011438 are acknowledged by S.F. JST-PRESTO Grant No. JPMJPR21B2 is acknowledged by S.K.

 \section*{AUTHOR CONTRIBUTIONS}
KYC and SF conceived and supervised the study. NSS developed the ASAP7 synthesis flow, ran SPICE simulations, and performed circuit-level experiments with sMTJs along with KK, QC and KS. KK, SK, SF and HO fabricated sMTJs. KK, QC, and SK ran the device-level sMTJ experiments. NAA, SN, TH have implemented the FPGA design for the learning and inference experiments. All authors have discussed the results and participated in writing and improving the manuscript. 
\section*{COMPETING INTERESTS}
The Authors declare no Competing Financial or Non-Financial Interests
\clearpage

\makeatletter
\def\bbordermatrix#1{\begingroup \m@th
  \@tempdima 4.75\p@
  \setbox\z@\vbox{%
    \def\cr{\crcr\noalign{\kern2\p@\global\let\cr\endline}}%
    \ialign{$##$\hfil\kern2\p@\kern\@tempdima&\thinspace\hfil$##$\hfil
      &&\quad\hfil$##$\hfil\crcr
      \omit\strut\hfil\crcr\noalign{\kern-\baselineskip}%
      #1\crcr\omit\strut\cr}}%
  \setbox\tw@\vbox{\unvcopy\z@\global\setbox\@ne\lastbox}%
  \setbox\tw@\hbox{\unhbox\@ne\unskip\global\setbox\@ne\lastbox}%
  \setbox\tw@\hbox{$\kern\wd\@ne\kern-\@tempdima\left[\kern-\wd\@ne
    \global\setbox\@ne\vbox{\box\@ne\kern2\p@}%
    \vcenter{\kern-\ht\@ne\unvbox\z@\kern-\baselineskip}\,\right]$}%
  \null\;\vbox{\kern\ht\@ne\box\tw@}\endgroup}
\makeatother

\setcounter{secnumdepth}{3}
\setlength{\belowcaptionskip}{1pt}

\tolerance=1
\emergencystretch=\maxdimen
\hyphenpenalty=10000
\hbadness=10000

%used to avoid whitespace before section headers
%\usepackage[compact]{titlesec}
\titlespacing{\section}{0pt}{*3}{*2}
\titlespacing{\subsection}{0pt}{*2}{*2}
\titlespacing{\subsubsection}{0pt}{*2}{*2}

% used to format (capitalize) Title
\titleformat{\section}{\filcenter\normalfont\small \bfseries}{\thesection.}{1em}{\MakeUppercase}

\onecolumngrid
\section*{Supplementary Information}
\setcounter{page}{1}
\beginsupplement

\section{Basic principles of probabilistic computing with hardware p-bits}
\label{sec:basic}
Probabilistic algorithms (e.g., sampling, inference, optimization) are performed with a network of p-bits interacting with each other \cite{camsari2019p}. The basic equation for the p-bit is given by: 
\begin{equation}
    m_i (t+\tau_N)=\text{sgn}\{\text{rand}(-1,1)+\text{tanh}[\beta I_i (t)]\},
    \label{eq:pbit}
\end{equation}
where $\text{rand(-1, 1)}$  represents a random number drawn from the uniform distribution in [$-$1,1], $\beta$ is the inverse algorithmic temperature, and  $I_i$ is the local field of p-bit ``$i$'' received from its neighbors. $\tau_N$ in this equation is defined as the neuron evaluation time \cite{faria2021hardware}. For the typical choice of 2-local (Ising-like) energy functions, $I_i$ is given by: 
\begin{equation}
    I_i(t+\tau_S)=\sum{J_{ij}m_j(t)}+h_i, 
    \label{eq:syn}
\end{equation}
where $J_{ij}$ are the weights and $h_i$ is the bias term for each individual p-bit. $\tau_S$ represents the synapse evaluation time. If the network of p-bits is symmetric ($J_{ij}=J_{ji}$), it is possible to define the following energy function: 
\begin{equation}
    E=-\left(\sum{J_{ij}m_im_j}+\sum{h_im_i}\right).
\end{equation}
In such a network of $N$ p-bits,  there are $2^N$ states, $S=\{1,2,\ldots,2^N\}$, and each state $j\in S$  is visited according to the Boltzmann-Gibbs distribution:
\begin{equation}
p_j =\frac{1}{Z}\exp(-\beta E_j).
\label{eq:boltz}
\end{equation}

The coupled evolution of Supplementary Eq.~\eqref{eq:pbit} and Supplementary Eq.~\eqref{eq:syn} represents a dynamical system and a wide variety of problems can be mapped onto this system, including combinatorial optimization, machine learning and quantum simulation problems, all phrased in terms of powerful physics-inspired Monte Carlo algorithms \cite{kaiser2021probabilistic,chowdhury2023full}. In this paper, we  focus on the two settings of probabilistic inference and Boltzmann learning. In either case, we will be interested in a fixed $\beta$ value (typically 1) and sample from the corresponding Boltzmann-Gibbs distribution of the network. We will show how the asynchronous and truly random bits generated by the sMTJs represent a low-level realization of physics-inspired probabilistic computation.

\section{Characterizing sMTJs through p-bits}
\label{sec:MTJ}
\begin{figure}[htbp]
    \centering
    \includegraphics[width=1\textwidth]{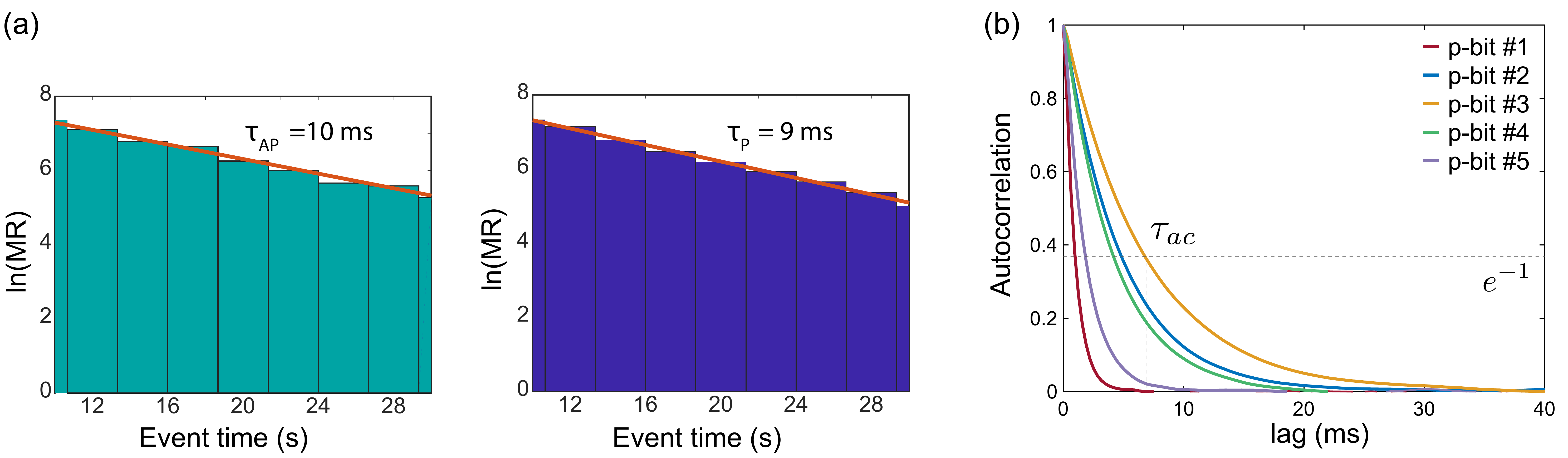}
    % \vspace{10pt} 
    \justify{\textbf{\ref{fig:FigS1}. sMTJ-based p-bit characterization}. (a) Histogram of  $\sf ln(MR)$ (number of magnetic relaxation (MR) events) as a function of the event times $ t_\text{event}$ for p-bit \#2, following the exponential distribution $1/\tau_{\text{P,AP}} \exp(-t_\text{event}/\tau_{\text{P,AP}})$, expected from a Poisson process (red lines). (b) Autocorrelation of sMTJ-based p-bits \#1-5. We define $\tau_{ac}$ as the time at which the normalized autocorrelation decays to $1/e$.}
     \refstepcounter{figure}\label{fig:FigS1}
 \end{figure}

\begin{figure}[t!]
    \centering \includegraphics[width=0.90\textwidth]{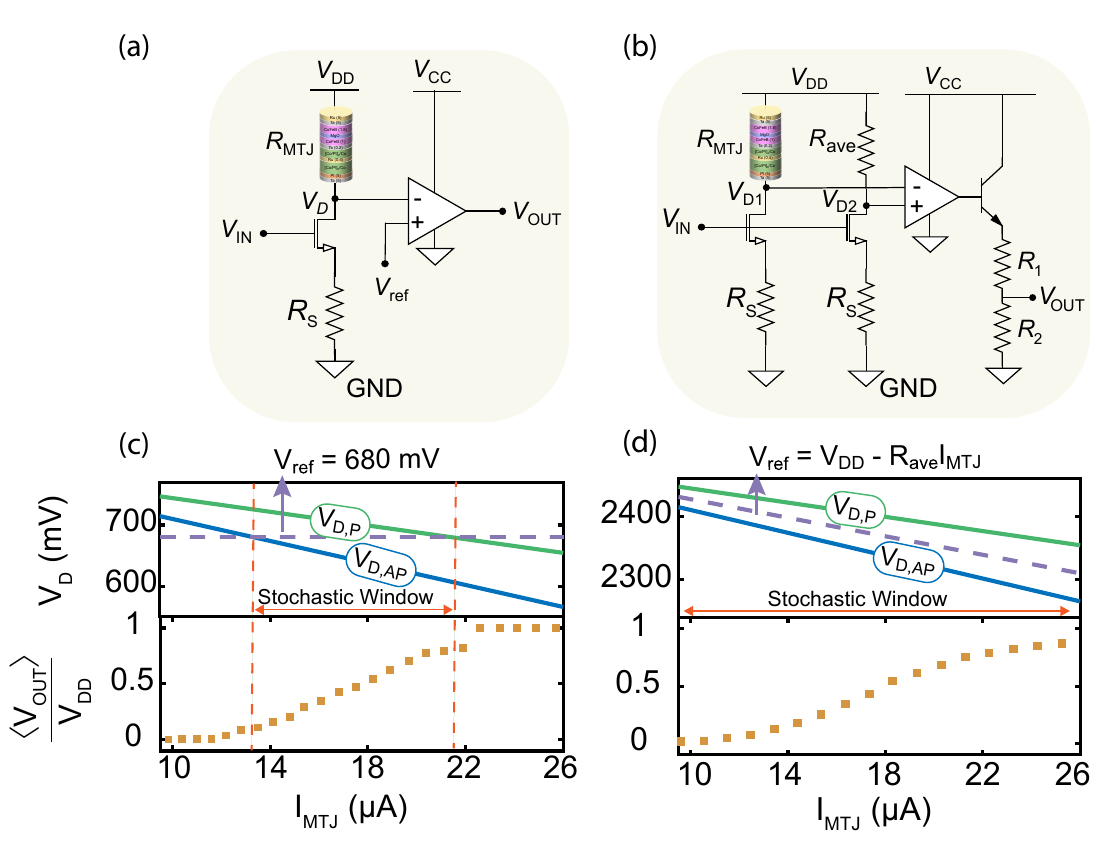}
    % \vspace{10pt} 
    \justify{\textbf{\ref{fig:FigS3}. sMTJ-based p-bit circuits} 
 (a) Single branch p-bit circuit diagram based on \cite{borders2019integer} with the following typical parameters:  $V_\text{DD} = 0.8 \ \si{V} $, $V_\text{CC} = 2 \ \si{V} $, $R_\text{S} = 10 \ \text{k}\Omega$, $V_\text{ref} = 680 \ \text{m}\si{V} $, and NMOS (2N7000) \cite{borders2019integer}  (b) Double branch p-bit circuit with a variable $V_\text{ref}$ and fixed $R_\text{S}$. $V_\text{DD} = 2.5 \ \si{V} $, $V_\text{CC} = 5 \ \si{V} $, $R_\text{S} = 47 \ \text{k}\Omega$, $R_\text{ave} = 1/2 \ (R_\text{P} + R_\text{AP})$, $R_1 = 100 \ \Omega$, $R_2 = 220 \ \Omega$, NMOS (ALD1101), and op-amp (ALD1702). (c) Single branch circuit of (a): the drain voltage as a function of current through the sMTJ, where $(V_{\text{D,AP}}=V_{\text{DD}}-R_{\text{AP}} I_{\text{MTJ}})$ and $(V_{\text{D,P}}=V_{\text{DD}}-R_{\text{P}} I_{\text{MTJ}})$. Vertical alignments show how the fixed reference voltage limits the stochastic window of the p-bit.  (d) Double branch circuit of (b): the drain voltage as a function of the sMTJ current, showing the variable reference voltage to the op-amp.} 
     \refstepcounter{figure}\label{fig:FigS3}
 \end{figure}

We first characterize the statistics of the sMTJs (in ~\ref{fig:FigS1}) by making measurements on the p-bit circuit described later in ~\ref{fig:FigS3}b. The fluctuations of the p-bit circuit are controlled entirely by the sMTJs. We  observe the outputs of 5 sMTJ-based p-bits and characterize their rate of fluctuations from event times and autocorrelations. 

In the main paper FIG.~\ref{fig:Fig1}g shows the fluctuations of p-bits \#1-5 at $I_{50/50}$, which is the current at which the sMTJs show 50/50 fluctuations  for the high- and low-resistance states.  The rate of fluctuations provides an estimate of the neuron evaluation time $\tau_N$ of Supplementary Eq.~\eqref{eq:pbit} after which the p-bit produces a new and independent random bit. 

\setcounter{table}{0}
\begin{table*}[h]
    \centering
     
\begin{tabular}{@{}>{\centering\arraybackslash}m{1.2cm}cc>{\centering\arraybackslash}m{1.8cm}c@{}}
\toprule
\multirow{2}{*}{\bf p-bit} & \multirow{2}{*}{\bfseries Mean MR} & \multirow{2}{*}{\bfseries Autocorr.} & \multirow{2}{*}{\bfseries TMR} &\multirow{2}{*}{\bfseries $I_{50/50}$} \\
& & & & \\
& {\bfseries time $(\tau)$} & {\bfseries time $(\tau_{ac})$} & & \\
\midrule
1 &2.4 ms& 0.97 ms& 64\% & 16 $\mu$A\\
2 &9.6 ms& 4.8 ms& 68\% & 19 $\mu$A\\
3 &14.4 ms&  6.8 ms& 65\% & 20 $\mu$A\\
4 &8.7 ms& 4.2 ms& 59\% & 17 $\mu$A\\
5 &4.2 ms&  1.8 ms& 65\% & 14 $\mu$A\\
\bottomrule

\end{tabular}

\justify{\textbf{\ref{tab:table1}. }{Table of results for all 5 p-bits reporting mean relaxation time $\tau=\sqrt{\tau_\text{P} \tau_{\text{AP}}}$, $\tau_{ac}$, TMR and $I_{50/50}$ of sMTJs. $I_{50/50}$ is defined as the absolute value of the current at which the perpendicular sMTJs show 50/50 fluctuations by canceling the uncompensated dipolar field resulting from the fixed layer.}}

 \refstepcounter{table}\label{tab:table1}
\end{table*}

 Even though the sMTJs were fabricated under the same conditions, slight differences in their volumes and shapes critically affect their relaxation times. The relaxation times $\tau_{\text{P,AP}}$ obeys a Néel-Arrhenius law \cite{coffey2012thermal}, described by $\tau=\tau_0\text{exp}(\Delta/k_B T)$, where $\tau_0$ is the attempt time and $\Delta$ is the energy barrier of the nanomagnet. In this paper, our magnets are not truly zero-barrier and their fluctuation rates exponentially depend on the energy barrier, $\Delta$, however, detailed theoretical analysis shows that this dependence is much less pronounced in low barrier nanomagnets \cite{kaiser2019subnanosecond,hassan2019low}. Moreover, as we experimentally show in Sections~3-4, p-bits in symmetric networks are agnostic to update orders \cite{faria2021hardware,aarts1989simulated,camsari2017stochasticL}, therefore variations in these time scales should not play a prohibitive role in scaled implementations of p-computers (see Supplementary Section~\ref{sec:randperm_supp} for another experiment showing this update order invariance).  

In ~\ref{fig:FigS1}b, we calculate the normalized autocorrelation of the p-bits using a 300 s time window with a sampling rate of 3.16 kHz, collecting $N\approx$ 949000 samples. The discrete autocorrelation function is defined as: 
\begin{equation}
C[m] = \frac{\displaystyle\sum_{n=0}^{N-1} V[n] V[n+m]}{\displaystyle\sum_{n=0}^{N-1} V[n]^2} ,
\end{equation}
 where $V[n]$ is the discrete sampled voltage read from the oscilloscope, and $m$ represents the discretized lag time. Since autocorrelations  typically show exponential decay \cite{kaiser2019subnanosecond}, we extract an autocorrelation time $\tau_{ac}$, by measuring and fitting the autocorrelation to a continuous function $C(t_{\text{lag}})=\text{exp}(-t_{\text{lag}}/{\tau}_\text{ac})$. As an additional measure to $\tau_{ac}$, we  also determine the frequency of magnetic relaxations that are characterized by an event time. The event time, $t_{\text{event}}$, is defined as the time between one event to the next. 
 Measuring the distribution of event times from parallel and antiparallel configurations, we observe that the p-bit outputs are distributed according to the exponential distribution, indicating a Poisson process \cite{funatsu2022local}. Overall, p-bits show a good degree of variation in the relaxation and autocorrelation times as well as in their TMR values and their 50/50 currents. These values are summarized in ~\ref{tab:table1}.

\section{A voltage-comparator based p-bit}
\label{sec:pcirc}
In this section, we describe a new p-bit circuit we designed in this work, comparing its characteristics to an earlier design implemented in \cite{borders2019integer}. ~\ref{fig:FigS3}a,b show both these designs. The earlier design was implemented in several small-scale experimental demonstrations using perpendicular sMTJs \cite{borders2019integer,pervaiz2019probabilistic,kaiser2022hardware}. In the original theoretical proposal \cite{camsari2017implementing}, however, circular or elliptical in-plane nanomagnets were used. In-plane low barrier magnets are very hard to pin, requiring spin polarized currents of  $\approx$100-500 $\mu$A or more for typical parameters \cite{hassan2019low}. If the magnetization provides continuous randomness providing all resistance values between $R_\text{P}$ to $R_\text{AP}$, this allows a faithful realization of Supplementary Eq.~\eqref{eq:pbit} as carefully discussed in \cite{hassan2021quant}. In such a case, spin-transfer-torque pinning is an unnecessary distraction, other than causing a read disturbance. Indeed, the fastest experimental p-bits are based on in-plane magnetic tunnel junctions \cite{camsari2021double,kobayashi2022external} 
 and variations of the design proposed in Ref.~\cite{camsari2017implementing} may still be useful in future implementations.

The experimental demonstrations including the present work have so far been primarily focused on perpendicular sMTJs to keep fluctuation speeds slow for practical reasons. Perpendicular sMTJs are easily pinned with spin currents around  $\approx$10 to $20 \ \mu$A for typical parameters \cite{hassan2021quant}. Unlike in-plane sMTJs, however, perpendicular sMTJs switch telegraphically and  they do not provide a uniform resistance between the two extremes. By a fortunate coincidence, the presence of the uncompensated dipolar field from the reference layer and easy pinning of perpendicular sMTJs appear to allow the realization of Supplementary Eq.~\eqref{eq:pbit} in hardware \cite{borders2019integer}, since the spin-torque pinning changes the 50/50 fluctuations of the sMTJ \cite{pervaiz2019probabilistic}. 

\begin{figure}[t!]
\includegraphics[width=0.875\linewidth]{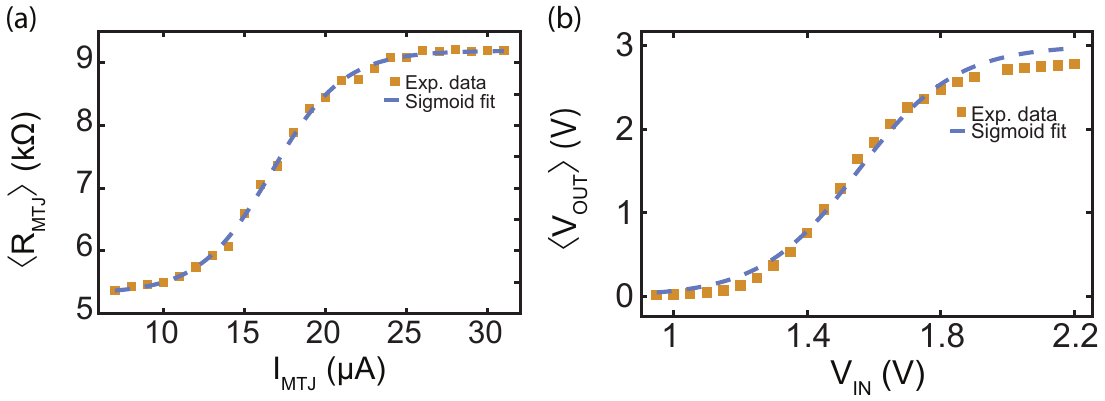}
    % \vspace{10pt} 
    \justify{\textbf{\ref{fig:pbitchar}. sMTJ-based p-bit circuit characterization} 
    (a) $\langle R_\text{MTJ} \rangle$, the time average (over a period of 3 minutes) of the sMTJ resistance, as a function of $I_\text{MTJ}$, the current flowing through the sMTJ. The yellow squares are experimental data, and the blue dashed line is a fit of the form $\langle R_\text{MTJ} \rangle = a + b/\{1 + \exp{[-c(I_\text{MTJ}-d)]}\}$, with $a=5.32 \ \text{k}\Omega$, $b=3.87 \ \text{k}\Omega$, $c=0.45 \ (\mu \si{A})^{-1}$, $d=16.6 \ \mu \si{A}$. (b) Experimentally measured $\langle V_\text{OUT} \rangle$, the time average (over a period of 3 minutes) of the output of the circuit shown in ~\ref{fig:FigS3}b, as a function of DC input voltage $V_\text{IN}$. The yellow squares are experimental data, and the blue dashed line is a fit of the form $\langle V_\text{OUT} \rangle = 1/2 \ V'_\text{CC} [\tanh[\beta(V_\text{IN} - V_0)] + 1]$, where $V_0 =1.55 \ \si{V}$, $\beta=3.43 \ \si{V}^{-1}$,  $V'_\text{CC} =3\ \si{V}$ is a reduced voltage from $V_{\text{CC}}=5\ \si{V}$.}
     \refstepcounter{figure}\label{fig:pbitchar}
 \end{figure}

 In the design shown in ~\ref{fig:FigS3}a, the comparator has a fixed reference voltage. This means that as a function of $V_\text{IN}$, the drain voltage swings between two extremes, $V_\text{D,AP}=V_{\text{DD}}-R_{\text{AP}} I_{\text{MTJ}}$ and $V_\text{D,P}=V_{\text{DD}}-R_\text{P} I_{\text{MTJ}}$. As shown in ~\ref{fig:FigS3}c, this limits the stochastic window of the p-bit. For the circuit shown in ~\ref{fig:FigS3}b, we have two parallel branches, one with the $R_{\text{MTJ}}$ and the other with $R_{\text{ave}}$, defined as $1/2(R_\text{P}$ + $R_\text{AP})$ of the sMTJ. The output nodes of these branches, taken from the drain of the transistor are compared by an operational amplifier. The key difference  as shown in ~\ref{fig:FigS3}d is that the voltage reference is variable in the circuit of ~\ref{fig:FigS3}b and a larger stochastic window can be obtained. Additionally, the differential nature of the voltage comparison in the new design allows the use of a single source resistance across all sMTJs, unlike the  circuit of ~\ref{fig:FigS3}a where the source resistance is adjusted individually \cite{borders2019integer}. Finally, the bipolar junction transistor at the output may not be necessary for integrated implementations, in this work, it simply functions as a buffer and lowers the output voltage to 3V, to safely interface with the FPGA.

\section{Experimentally designing the voltage-comparator based p-bit}
\label{sec:expcirc}
Although the circuit shown in ~\ref{fig:FigS3}b is operated at only 50/50 fluctuations to provide asynchronous clocks, this circuit can also be used as a full p-bit whose output probability is tunable by the input voltage $V_\text{{IN}}$.

Each of the sMTJs used in this work has different characteristics, as such $R_\text{ave}$ and $V_\text{IN}$ should be customized for the specific sMTJ used in a p-bit circuit. We first characterize the specific sMTJ by experimentally measuring the $\langle R_\text{MTJ} \rangle$ - $I_\text{MTJ}$ (~\ref{fig:pbitchar}a), the value of $R_\text{P}$, and the value of $R_\text{AP}$. The value of the $R_\text{ave}$ resistor is given by $R_\text{ave} = 1/2 \ (R_\text{P} + R_\text{AP})$. 

As shown in ~\ref{fig:FigS3}b, the NMOS transistor and the $R_\text{S}$ in either branch of the circuit form a constant current source. The two current sources in the two branches form a current mirror, so they have the same I-V characteristics. Note that the I-V characteristics of the circuit (with ALD1101 NMOS transistors and $R_\text{S} = 47 \ \text{k} \Omega$) do not depend on the $R_\text{P}$ and $R_\text{AP}$ values of the specific sMTJ used.

The ALD1101 current source can be characterized using the following setup: we replace the sMTJ and the $R_\text{ave}$ in the left and right branches of the circuit with two $10 \ \text{k}\Omega$ resistors, we connect $V_\text{DD}$ to a 2.5 V power supply, and we leave $V_\text{CC}$ unconnected. ~\ref{table:tableS2} shows the experimentally measured I-V relationship.

\begin{table}[!ht]
\footnotesize
{\textbf{\ref{table:tableS2}.} ALD1101 constant current source characterization: experimentally measured $I_\text{NMOS}$ - $V_\text{IN}$ relationship. The ALD1101 constant current source consists of the ALD1101 matched pair NMOS (the ALD1101 is a single IC consisting of two NMOS's with their threshhold voltages matched to within $10$ mV of difference) and the two $R_\text{S}$ resistors, as shown in ~\ref{fig:FigS3}b.}
\resizebox{\textwidth}{!}{%
\begin{tabular}{|c|c|c|c|c|c|c|c|c|c|c|c|c|c|c|c|c|c|c|c|c|}\hline
$V_\text{IN}$ (mV) & 600 & 700 & 800 & 900 & 1000 & 1100   & 1200 & 1300 & 1400 & 1500 & 1600 & 1700 & 1800 & 1900 & 2000 & 2100 & 2200 & 2300 & 2400\\ \hline
$I_\text{NMOS}$ ($\mu \si{A}$) & 0.974 & 2.24 & 3.83 & 5.55 & 7.37 & 9.21 & 11.1 & 13.0 & 15.0 & 16.9 & 18.9 & 20.9 & 22.8 & 24.8 & 26.8 & 28.8 & 30.8 & 32.8 & 34.8\\
\hline
\end{tabular}}
\caption*{}
\refstepcounter{table}\label{table:tableS2}
\end{table}\vspace{-20pt}
We discovered that the variation in the $V_\text{TH}$ of each ALD1101 IC is important. While the shape of the $I_\text{NMOS}$ - $V_\text{IN}$ relationship in ~\ref{table:tableS2} is the same for each ALD1101 IC, the $V_\text{TH}$ variations lead to offsets, shifting the $I_\text{NMOS}$ - $V_\text{IN}$ curves up or down. Remeasuring of all the data points in ~\ref{table:tableS2} is not required, only one measurement is needed: using the same characterization setup described above, we find and record the $V_\text{IN}$ that produces $I_\text{NMOS} = 16.9 \ \mu \si{A}$. The difference of this $V_\text{IN}$ with the value recorded in ~\ref{table:tableS2} (1500 mV) gives the offset to the entire $I_\text{NMOS}$ - $V_\text{IN}$ curve of the specific ALD1101 used. We simply add this offset to all data points in the reference $I_\text{NMOS}$ - $V_\text{IN}$ curve. We use the experimentally measured $\langle R_\text{MTJ} \rangle$ - $I_\text{MTJ}$ curve of the specific sMTJ and the $I_\text{NMOS}$ - $V_\text{IN}$ curve of the specific ALD1101 IC to determine the stochastic range of $V_\text{IN}$.  

\section{FPGA Architecture}
\label{sec:FPGA}
\subsection{p-bit and MAC Unit} To evaluate the performance of the random number generators, we first implemented fully digital probabilistic bits (p-bits) on a Kintex UltraScale KU040 FPGA Development Board. A single p-bit consists of a tanh lookup table (LUT), a random number generator (RNG) and a comparator to implement Supplementary Eq.~\eqref{eq:pbit}. In this work, we used 32-bit LUTs and RNGs. There is also a multiplier–accumulator (MAC) unit to compute Supplementary Eq.~\eqref{eq:syn} from the neighbor p-bits and provide the input signal for the LUT. The p-bits are interconnected in a fixed hardware topology where the weights and biases, $J_{ij}, h_i$ are stored in 10-bit registers and a digital multiply-accumulate operation with $m_j \in \{0,1\}$ selects a particular weight or not. In the FPGA, we switch from a bipolar p-bit $m_i\in \{-1,1\}$ to a binary formulation $m_i\in \{0,1\}$ \cite{aadit2022massively}. In our hybrid CMOS + sMTJ based p-bits, the architecture remains the same except the sMTJs serve as the clocks for the LFSRs and the p-bits. 

\subsection{RNG Unit} As shown in Algorithm 1 in the Supplementary information, we used three different types of RNGs: linear feedback shift register (LFSR), Xoshiro \cite{blackman2021scrambled} and sMTJ-driven LFSR to compare the quality of randomness. We used 32-bit RNGs and compared the RNG outputs with the 32-bit LUTs to implement Supplementary Eq.~\eqref{eq:pbit}. LFSR involves a linear shift operation on all the bits and an XNOR operation on some bits based on the selected taps. Unique seeds were used for each RNG and unique taps were used for RNGs in each p-bit block while ensuring maximal-length outputs. In contrast, a Xoshiro RNG involves linear shift and rotation operations on 32-bit words as well as XORing between subsequent words. In the all-digital versions, LFSR and Xoshiro RNGs are driven by digital clocks. However, in the hybrid design where sMTJs serve as the clocks for the RNGs, sMTJ + LFSR is considered as a new RNG unit. Using sMTJ-based p-bit removes the need for the 32-bit RNG and the 32-bit LUT.  

\subsection{Clocking and Sampling Unit}
For the fully digital CMOS p-bits, each p-bit and its RNG is driven by system clocks generated on the low-voltage differential signaling (LVDS) clock-capable pins of the FPGA. These clocks are generated using Xilinx LogiCORE™ IP clocking wizard and  mixed-mode clock manager (MMCM) module. Each of the five clocks operates at 15 MHz with shifted phases and is highly accurate with low jitter noise. In the Boltzmann learning example and for the NIST tests, to make these clocks comparable to sMTJs, we used frequency divider circuits to slow down the clocks to $\approx2$ kHz.  For the sMTJ-driven p-bits, sMTJs replace the FPGA clocks and drive the p-bits and the RNGs externally using the peripheral module (PMOD) interface. FPGA samples the sMTJ outputs with a fast system clock of 75 MHz. 

\subsection{Data Programming and Acquisition Unit}
We used MATLAB 2022a as the host program to read-write data to and from the FPGA through the USB-JTAG interface. MATLAB communicates with the FPGA board via AXI4 (Advanced eXtensible Interface 4) protocol where MATLAB works as the AXI master to drive a slave memory-mapped registered bank and Block RAMs (BRAM) inside the FPGA. We used airHDL \cite{airHDL}, a memory management tool to assign the memory addresses for the register bank and the BRAMs. The weights $J$, $h$ of the full adder circuit are programmed through MATLAB. In the inference and the Boltzmann learning example, the p-bit outputs were read from MATLAB as batches of sweeps that were initially sampled at 2kHz and stored in a BRAM. For the NIST tests, however, we sampled and stored all the sweeps in the BRAM at a much higher frequency ($\approx$10 kHz) compared to the clock frequency of $\approx2$ kHz and then downsampled the data to a designated frequency. This procedure ensured that we did not lose any samples. MATLAB reads the BRAM data in burst mode and performs the downsampling.

\section{Inference on the probabilistic full adder}
\label{sec:samplingsupp}

%abc:ignore
Inside the FPGA, we construct digital p-bits that behave according to Supplementary Eq.~\eqref{eq:pbit} and interconnections between p-bits that behave according to Supplementary Eq.~\eqref{eq:syn}. Each p-bit has a PRNG, a LUT for the hyperbolic tangent function and 10-bit weights in fixed-precision, 1 sign, 6 integer, and 3 fractional bits (s[6][3]). 
Supplementary Eq.~\eqref{eq:syn} is implemented by  a multiply-accumulate unit inside the FPGA, whose multiplication reduces to simple multiplexing since a given weight $J_{ij}$ is either taken or not if $m_j$ is 0 or 1. 
An important consideration in ensuring the p-bit network reaches the equilibrium is the necessity of fast synapse times ($\tau_s$) compared to neuron times $\tau_n$ \cite{faria2021hardware}. In our context this requirement ($\tau_s < \tau_n$) is naturally satisfied because the combinational logic inside the FPGA, which computes Supplementary Eq.~\eqref{eq:syn} with about 10 ns delays is orders of magnitude faster than both our deliberately slowed digital clocks and our sMTJs. In scaled and  integrated implementations with fast p-bits with GHz fluctuations, this necessity requires careful design. In the case sMTJ clocks, there is also the theoretical possibility of parallel updates by simultaneously switching sMTJs. Practically this is not a concern due to the extremely low probability of such an event which would be washed over thousands of samples anyway. Our results with sMTJs in FIG.~\ref{fig:Fig2}c,d and in FIG.~\ref{fig:Fig3}c,d indicate that the sMTJs reproduce the ideal distributions well.  

A full block diagram of the FPGA unit is shown in FIG.~\ref{fig:Fig3}a. To rule out any spurious correlations, the starting states (seeds) used for the LFSR and Xoshiro are randomized for each p-bit. LFSRs also use unique sets of random taps while ensuring maximum-length outputs. 

 In this setting, for each RNG, we cumulatively sampled $10^6$ states from the p-bits starting from a random initial state. A system state out of 5-p-bits can be defined from 0 to $2^N-1$ such that  state 0 is $p_1p_2p_3p_4p_5 = 00000$ and state 31 is $p_1p_2p_3p_4p_5 = 11111$. We define the single update of each p-bit according to Supplementary Eq.~\eqref{eq:pbit}-\eqref{eq:syn} as a sweep. Due to their digital nature, defining exact times to perform a sweep for LFSR and Xoshiro is straightforward. With a driving clock frequency of 2 kHz, we perform one sweep and then record it as a new state.  However, sampling states from sMTJ-clocked LFSR p-bits is not straightforward due to the analog nature of fluctuations of the sMTJs. The relaxation time of the slowest sMTJ is $\approx$20 ms (50Hz) that is 40$\times$ slower than the sampling frequency of 2 kHz. For this reason, we collect 40$\times$ more data points from sMTJ-based p-bits and downsampled them to obtain independent samples. The oversampling and subsequent downsampling of sMTJ data is due to our 40$\times$ faster readout process, not an inherent sMTJ limitation, and is implemented to use the common sampler for each RNG setup. The sampling frequency can easily be adjusted, removing the need for oversampling and downsampling.

\SetKwComment{Comment}{\normalfont $\triangleright$ }{}
\SetKwInOut{Input}{Input}
\SetKwInOut{Output}{Output}
\SetKwInput{KwSampler}{RNG}
\begin{algorithm}
\caption{Learning the full adder on Deep BMs}
\label{alg:alg2}
\Input{number of samples $N$,  number of truth table lines $T$, epochs $N_{\text{L}}$, learning rate $\varepsilon$, regularization $\lambda$}
\KwSampler{LFSR, Xoshiro, sMTJ-clocked LFSR}
\Output{learned weights $J_{\text{new}}$ and biases $h_{\text{new}}$}
$J_{\text{new}} \gets 0.01*\text{randi}([-1,1])$\;
$h_{\text{new}} \gets \text{randi}([-1,1])$\;
\For{$i\gets 1$ \KwTo $N_{\text{L}}$}{
   % \For{$j \gets 1 $ \KwTo $N_{\text{B}}$}{
        $J_{\text{FPGA}} \gets J_{\text{new}}$\;
        \tcc{positive phase}
        \For{$j \gets 1$ \KwTo $T$}{            
            $h_{\text{$T$}} \gets \text{ clamping to truth table values}$\;
            $h_{\text{FPGA}} \gets h_{\text{$T$}}+h_{\text{new}}$\;
            
            $\{r\} \gets \text {RNG}() $\;
            
            $\{m\} \gets \text{FPGA}(N, \{r\})$\;
         }
         $\langle m_im_j\rangle_{\text{data}} = \{m\}\{m\}^{\text{T}}$\;
        \tcc{negative phase}
        $h_{\text{FPGA}} \gets h_{\text{new}}$\;
        $\{r\} \gets \text {RNG}() $\;
        $\{m\} \gets \text{FPGA}(N\times T, \{r\})$ \; 
        $\langle m_im_j\rangle_{\text{model}} = \{m\}\{m\}^{\text{T}}$\; 
        \tcc{update weights and biases}
        $J_{\text{new}} \gets J_{\text{new}} + \Delta J_{ij}$\;  
        $h_{\text{new}} \gets h_{\text{new}} +\Delta h_i$\; 
    }

\end{algorithm}

 \section{Learning the full adder on the 32-node chimera lattice}
 \label{sec:learningsupp}

 Boltzmann machines can be trained using the contrastive divergence algorithm. There are two phases during the training of Boltzmann machines as shown in FIG.~\ref{fig:Fig3}a and Algorithm \ref{alg:alg2}. The first one is the positive phase where the network operates in its clamped condition under the direct influence of the training samples. The next one is the negative phase when the network is allowed to run freely without having any environmental input. The update rules can be obtained by minimizing the KL divergence between the data and the model distributions \cite{ackley1985learning}: 
\begin{eqnarray}
    \Delta J_{ij}=\varepsilon\bigg(\langle m_im_j\rangle_\text{{data}}-\langle m_im_j\rangle_\text{{model}}-\lambda J_{ij}\bigg) \label{eq:delta_J} \\
    \Delta h_{i} = \varepsilon\bigg(\langle m_i\rangle_\text{{data}}-\langle m_i\rangle_\text{{model}}-\lambda h_{i}\bigg),
   \label{eq:delta_h}
\end{eqnarray}
where $\epsilon$ is the learning rate, $\lambda$ is the regularization parameter, $\langle m_im_j\rangle_\text{{data}}$ is the average correlation between two neurons in the positive phase, and $\langle m_im_j\rangle_\text{{model}}$ is the average correlation between two neurons in the negative phase.

As described in the previous section, five sMTJ-based p-bits are used here as clocks for digital p-bits with 32-bit LFSRs in FPGA and each LFSR starts with a unique seed. We also used 5 different sets of taps for the LFSRs, ensuring maximal-length output.
The $\langle m_im_j\rangle_\text{{data}}$ is obtained by clamping visible bits to the eight lines in the truth table of the full adder (FIG.~\ref{fig:Fig2}b). Then the $\langle m_im_j\rangle_\text{{model}}$ is calculated in the negative phase where the clamp is removed. We obtain the updated weights and biases using Supplementary Eqs.~\eqref{eq:delta_J},\eqref{eq:delta_h} and repeat the weight learning for 500 epochs. We also naturally adopt the persistent contrastive divergence (PCD) algorithm that runs a continuous Markov Chain from the last state of the previous update to the next \cite{10.1016/RBM}. This is because the FPGA holds the previous state of the chain and continues sampling from that state as new weights are loaded. The hyperparameters  we used while learning are as follows:  inverse temperature $\beta$ = 1, learning rate $\varepsilon$ = 0.003, and regularization $\lambda$ = 0.005.

%abc:ignore
 The 32-node Chimera graph is bipartite. This means that p-bits in a Chimera topology can be updated in parallel in two blocks. In order to make our system resemble eventual, fully-asynchronous systems, we distributed our 5 available sMTJ clocks over these two blocks, ensuring that no sMTJ clock serviced two p-bits of the same block to avoid parallel updates (FIG.~\ref{fig:Fig3}b bottom panel shows our clock distribution over the p-bits). For the fully digital setup, we distributed the digital clocks similarly. As in the inference experiments in the previous section, LFSR and Xoshiro RNGs were driven at 2 kHz. We used the same 5 sMTJ-based p-bits we characterized (\ref{fig:FigS1}) and sampled them at 2 kHz. To train the full adder on the LFSR and Xoshiro RNG-based p-bit network, we took 400 sweeps per epoch for a total of 500 epochs. For the sMTJ-clocked LFSR based p-bit network, we took 16000 sweeps per epoch and a total of 500 epochs. We took 40$\times$ more sweeps for the sMTJ because they were sampled at 2 kHz (40$\times$ faster than their autocorrelation) in order to produce the same number of independent sweeps for all solvers. 
%abc:endignore

\begin{figure}[htbp]
    \centering
    \includegraphics[width=1\textwidth]{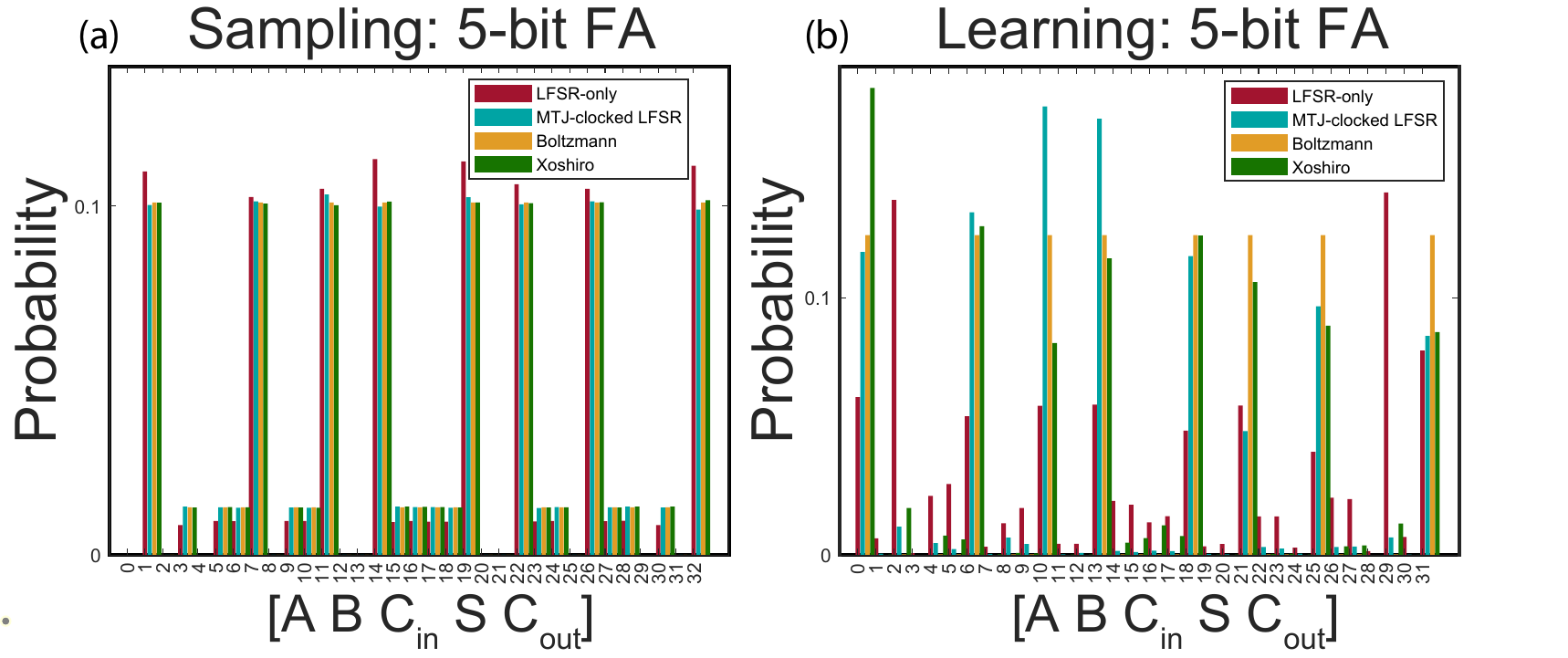}
    % \vspace{10pt} 
    \justify{\textbf{\ref{fig:FigS4}. }Extended data for probabilistic sampling and learning experiments, showing all 32 states of the 5-bit full adder for LFSR and sMTJ-clocked LFSR. (a) Sampling data is taken at $10^6$ sweeps, (b) Learning data is taken at epoch = 400 where the number of sweeps per epoch = 400 for LFSR-only and the number of sweeps per epoch = 16000 for the sMTJ-clocked LFSR.
    (c) Sampling data for Xoshiro (d) Learning data for Xoshiro}
     \refstepcounter{figure}\label{fig:FigS4}
     \vspace{-15pt} 
 \end{figure}

In ~\ref{fig:FigS4} we provide the full histograms (32-states) of the full adder for the sampling and learning experiments. Only parts of the histograms are shown in the main text for clarity. In ~\ref{fig:Fig_seed_tap_bias} we show that this bias is not random and never goes away, rather it is a consistent systematic bias by starting the LFSRs at different uniform random initial conditions (seeds) and different maximum-length taps.

\section{Update order invariance of Boltzmann Networks}
 \label{sec:randperm_supp}

\begin{figure}[htbp]
    \centering
    \includegraphics[width=0.75\textwidth]{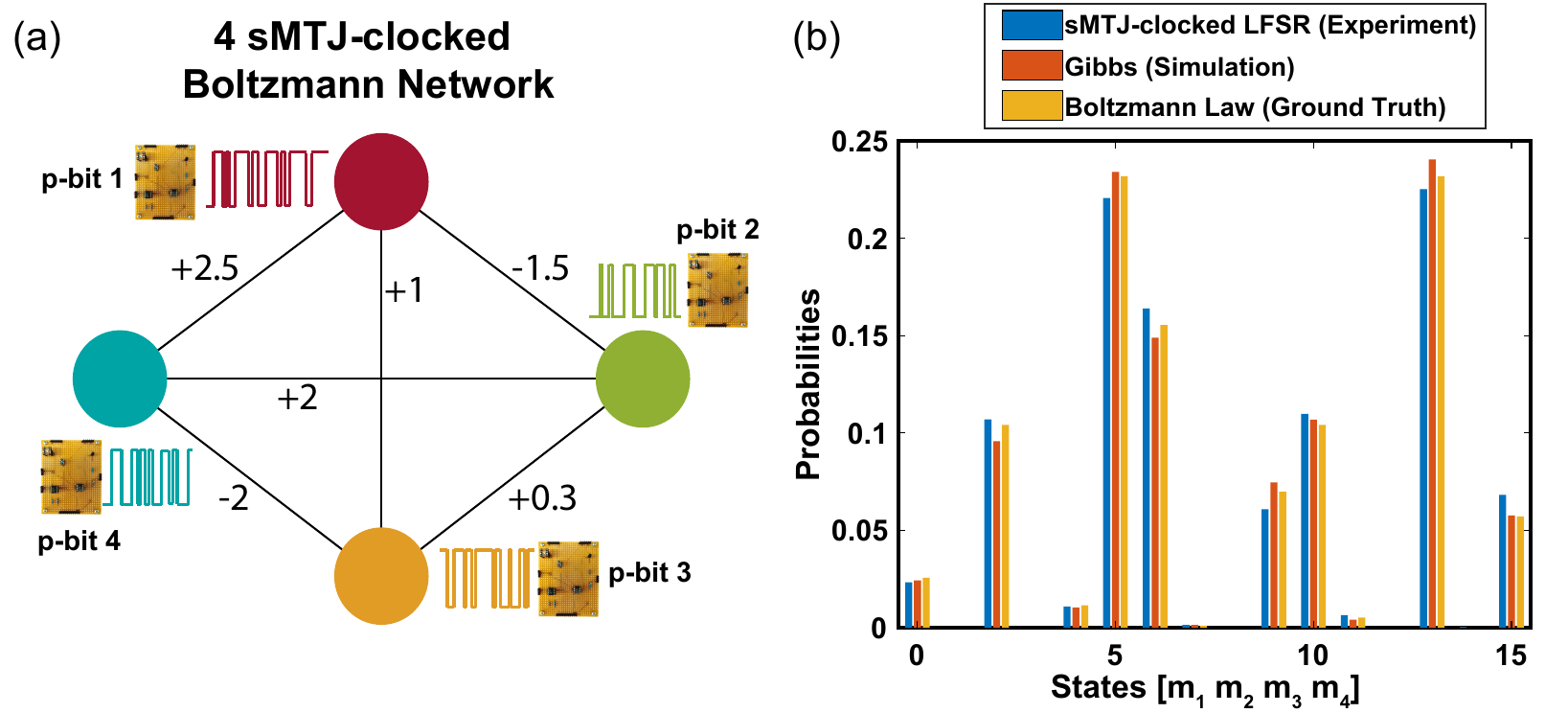}
    % \vspace{10pt}    
    \justify{\textbf{\ref{fig:Fig_randperm}. }(a) Fully connected undirected Boltzmann network with 4 nodes. Different sMTJs drive the LFSRs in the p-bits corresponding to each node. (b) Experimental results of 4 sMTJ-clocked LFSR-based p-bits (1e4 iterations) performing probabilistic inference. This asynchronous clocking of LFSR-based p-bits matched the random Gibbs Sampling simulation results (1e4 iterations), which also matched the Boltzmann Law ground truth for these undirected networks. Note that the Gibbs simulation randomized the update order of the network, choosing one of 4!=24 possible permutations randomly at each iteration. The weight matrix is shown on the nodal interconnections, and the bias was 0 for p-bit 1, 3 and 4, but set to 0.3 for p-bit 2.}
     \refstepcounter{figure}\label{fig:Fig_randperm}
     \vspace{-15pt} 
 \end{figure}

 In ~\ref{fig:Fig_randperm}, we perform additional experiments to study how undirected Boltzmann networks are invariant to update orders in the system, significantly easing experimental difficulties in scaled-out implementations of p-bit networks. In this experiment, a fully connected undirected p-bit network had nodes clocked asynchronously with independent sMTJs. 
Even with only $10^4$ samples, a very close  match is seen between the distributions corresponding to sMTJ-clocked LFSRs and Gibbs sampling in simulation. It is important to note that Gibbs sampling was performed with \textit{randperm} updates, where a new update order (out of 4!=24 possibilities) was sampled (without replacement) at each iteration. Despite this highly random updating scheme, both Gibbs (ideal simulation) and the sMTJ-clocked system eventually reach the ground truth, represented by the Boltzmann distribution. This remarkable feature of update invariance in Boltzmann networks \cite{aarts1989simulated} significantly eases fabrication difficulties in eventual sMTJ-based large-scale p-bit networks. Going further, the investigation of ``time-to-equilibrium'' that determines the model averages and correlations in Algorithm~\ref{alg:alg2} remains to be one of the future challenges of the field. 

 \section{Sampling from arbitrary distributions}
  \label{sec:arbitrary}
\begin{figure*}[b!]
    \centering
    \includegraphics[width=0.825\textwidth]{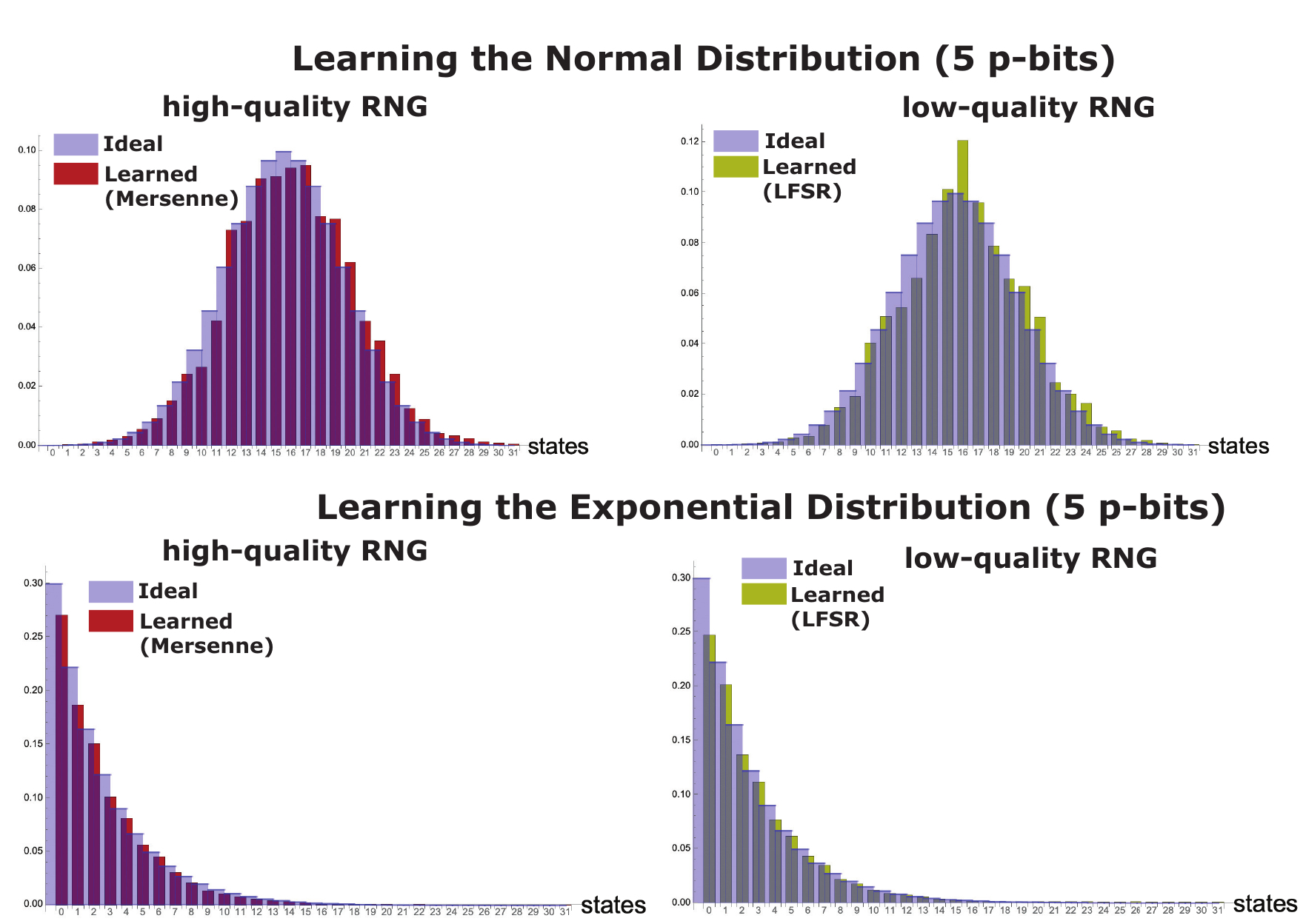}
    \justify{\textbf{\ref{fig:reb1}. }Networks of p-bits can be connected to sample from arbitrary probability distributions: on the left of the top panel is a network of 5 p-bits approximating a normal distribution where training and inference are both done with high-quality Mersenne Twister-based PRNGs. The blue-filled lines are the exact probability density functions (PDF), discretely sampled in both figures. Bottom panel shows a similar result to learn the exponential distribution.
    Histograms show the inference on a 5 p-bit network that has been trained using the contrastive divergence algorithm. On the right is the same result (inference followed by learning) using 5 p-bits using LFSRs only (32-bit with random taps and initial conditions). All histograms are obtained with 50,000 samples. All models are trained on a 5-bit all-to-all fully visible network (weights and biases not shown).}\vspace{-15pt}
     \refstepcounter{figure}\label{fig:reb1}
\end{figure*}

 In this section, we show how the CD algorithm can be used to create networks of p-bits that can sample from arbitrary distributions up to M-bit accuracy. We start by defining a PDF for a given distribution. As an example, we choose a normal distribution with $\mu=16$ and $\sigma=4$ to be approximated by a $N=5$ p-bit network with $2^N=32$ states. Once the PDF is specified, one approach is to create a truth table matrix of size $V=NT\times N$, where the rows of the truth table are repeated according to the specified PDF. Assuming a fully visible network, we then calculate the data correlations from the truth table, $V^\dagger V$. We then perform the standard contrastive divergence algorithm. ~\ref{fig:reb1} shows inference and learning of a normal and exponential distribution with Mersenne Twister-based high-quality RNGs and LFSR-based low-quality RNGs. Even though the Mersenne-based learning looks marginally better, we do not see a strong LFSR bias in such low-dimensional learning tasks. Next, we test the difference between low and high-quality RNGs in much larger networks including MNIST handwritten digits.

\section{Training MNIST with deep Boltzmann machine: LFSR vs Xoshiro}
 \label{sec:trainDBM_supp}

\begin{figure}[t!]
    \centering
    \includegraphics[width=0.5\textwidth]{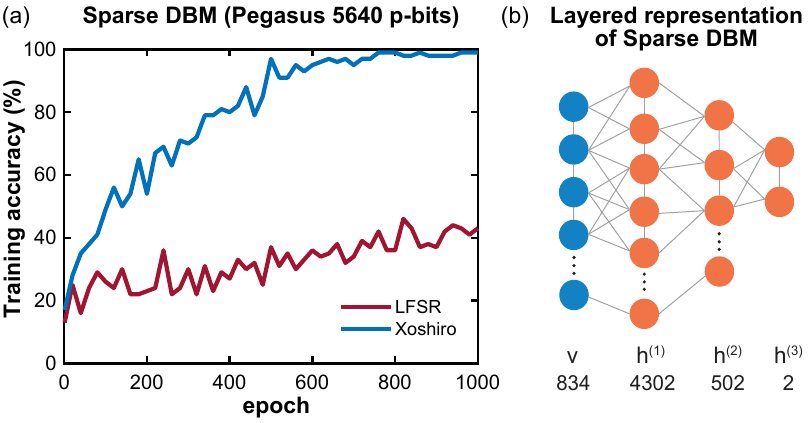}
    % \vspace{10pt} 
    \justify{\textbf{\ref{fig:Fig_mnist}. }(a) Training accuracy of 100 handwritten digits from MNIST with a sparse Deep Boltzmann Machine (Pegasus 5640 p-bits) for 1000 epochs using both LFSR and Xoshiro. This shows a significant discrepancy in training accuracy depending on whether p-bits are implemented with LFSR or Xoshiro. Despite being identical excluding the RNG implementation, the former does not suffice due to the low accuracy at around 40\% even after 1000 epochs, the latter reaches 100\% accuracy with the same number of epochs and the same set of hyper-parameters. (b) Illustration of sparse DBM (Pegasus 5640 p-bits) with 1 layer of visible units and 3 layers of hidden units where both the inter-layer and intra-layer connections are allowed. }
     \refstepcounter{figure}\label{fig:Fig_mnist}
     \vspace{-15pt} 
 \end{figure}

We now present a more complex task of training a subset of MNIST handwritten digits \cite{lecun1998mnist} using Algorithm \ref{alg:alg2} that also suffers from the low-quality randomness of LFSR. Pegasus 5640 p-bits \cite{dattani2019pegasus} is chosen as the sparse DBM where we randomly distribute the visible and hidden units which yield multiple hidden layers as shown in ~\ref{fig:Fig_mnist}b. The details of this method are described in \cite{niazi2023training}. Here, we choose a subset of MNIST data including 100 images with no down-sampling and train them in 10 mini batches using the CD algorithm for both LFSR and Xoshiro. All the hyperparameters remain the same for these two cases which are as follows:  inverse temperature $\beta$ = 1, learning rate $\varepsilon$ varies linearly from 0.03 to 0.003 for 1000 epochs.

 ~\ref{fig:Fig_mnist}a shows the comparison of training accuracy of 100 MNIST digits between the implementation of LFSR and Xoshiro. While LFSR-based training barely reaches 40\% accuracy in 1000 epochs, the Xoshiro-based training goes to 100\% strongly indicating how the quality of randomness matters in terms of training such a large network.

\section{Analyzing LFSR Bias over different taps and initial conditions }
 \label{sec:bias}

\begin{figure}[htbp]
    \centering
    \includegraphics[width=0.95\textwidth]{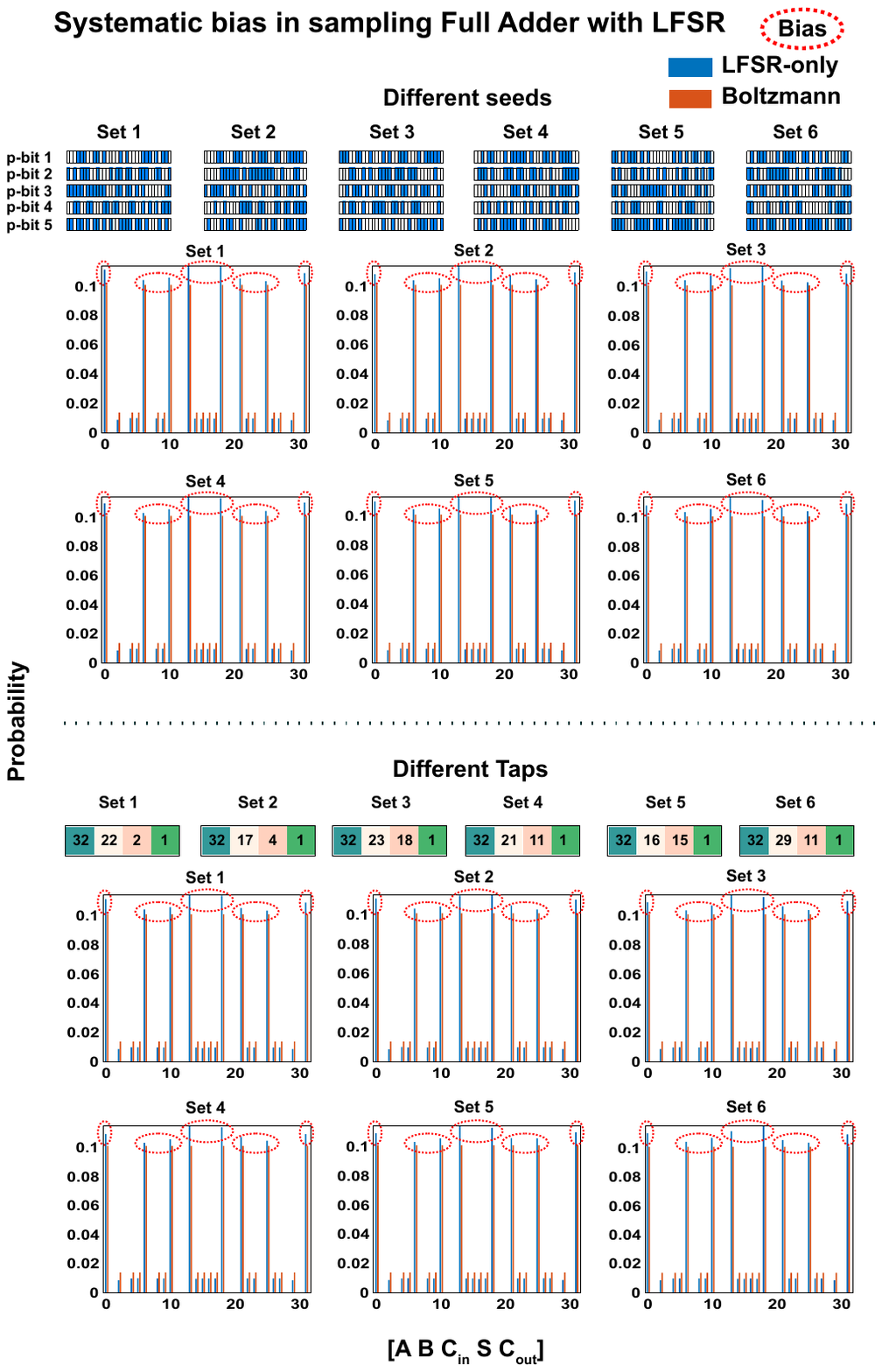}
    % \vspace{10pt} 
    \justify{\textbf{\ref{fig:Fig_seed_tap_bias}. }Demonstration of consistent systematic bias in sampling the 5 p-bit full adder using multiple sets of seeds and taps for the LFSR. The seeds are chosen uniform randomly and the taps are chosen to provide the maximum length of the LFSR. Irrespective of the seeds and taps, the bias never goes away, indicating a systematic bias.}
     \refstepcounter{figure}\label{fig:Fig_seed_tap_bias}
     % \vspace{-15pt} 
 \end{figure}

~\ref{fig:Fig_seed_tap_bias} presents a detailed analysis of LFSR as a random number generator. LFSR RNG quality is measured by its ability to perform bias-free sampling on the probabilistic full adder. Across different seeds and taps, even for $10^6$ iterations, a consistent bias is always seen as demonstrated by the variations in histogram peaks.

\section{NIST tests on Xoshiro, LFSR and sMTJ clocked LFSRs }
\label{sec:NIST}
NIST tests are widely used to evaluate the quality of randomness, so we conducted standard randomness tests on the bitstreams generated by the LFSR, Xoshiro, and LFSR+sMTJ using the NIST Statistical Test Suite \cite{Rukhin2010nist}. We applied all 16 different NIST tests to the bitstreams. For LFSR and Xoshiro, we used MATLAB to generate bitstreams since given taps and initial conditions these bitstreams are fully reproducible. 

We used a modified experimental setup to obtain the bitstreams for LFSR+sMTJ. First, the inverse temperature value $\beta$ in Eq. (1) was set to 0 to get 50/50 fluctuations. Second, we sampled bitstreams of 650000 bits with a $10$ kHz sampling rate in BRAM blocks of the FPGA. Then, we downsampled the bitstream by a factor $k$. This is equivalent to sampling the bitstream with frequency $f_{k}$ that $f_{k} = 10000 \ \mathrm{Hz} / k$. The length of the bitstream then becomes $650000 / k$. The reason for this downsampling is to obtain independent samples from the sMTJs whose fluctuations times are far above 100 microseconds. 

We used p-bit \#3 to generate the bitstreams for LFSR+sMTJ. The result ‘Random’ represents that the bitstream passes the tests whereas ‘Non-Random’ means the bitstream fails. \ref{table:tableS1} summarizes the results of the NIST tests for the LFSR+sMTJ with $k=601$ and $k=2001$, Xoshiro and LFSR. The results without oversampling issue show the good quality of randomness generated by LFSR+sMTJ when k=601 or $f_k=16.63$ Hz, corresponding to a period of 60 milliseconds, of the same order of our sMTJ fluctuations reported in ~\ref{tab:table1}. The result of $k=2001$ fails Maurer’s Universal Test (Test \#9) because the downsampled bitstream with length $650,000 / 2001 \approx 324$ is not long enough for applying that test. Corroborating our results in the main text, LFSR+sMTJ and Xoshiro pass all the tests, while LFSR fails one test. \\

\vspace{-10pt} 
\begin{table}[!ht]
    \centering
    \textbf{\ref{table:tableS1}. }Results of tests on bit streams specified in standard NIST SP800-22a
    \setlength{\tabcolsep}{1mm}
    \begin{tabular}{ccccccc}
    \hline \hline
Test \# & Test Name                  & \begin{tabular}[c]{@{}l@{}}Sub-tests\end{tabular} & \begin{tabular}[c]{@{}l@{}}LFSR + sMTJ \\$k = 601$\\$f_{k} = 16.63 \ \mathrm{Hz}$\end{tabular} & \begin{tabular}[c]{@{}l@{}}LFSR+sMTJ \\$k = 2001$\\$f_{k} = 5.00 \ \mathrm{Hz}$\end{tabular} & \begin{tabular}[c]{@{}l@{}}Xoshiro\end{tabular} & \begin{tabular}[c]{@{}l@{}}LFSR\end{tabular}  \\ \hline
1       & Frequency                  & 1                   & ~Random           & ~Random         & ~Random     & ~Random      \\
2       & Frequency within a Block   & 1                   & ~Random           & ~Random         & ~Random     & ~Random      \\
3       & Runs                       & 1                   & ~Random           & ~Random         & ~Random     & ~Random      \\
4       & Longest run of ones        & 1                   & ~Random           & ~Random         & ~Random     & ~Random      \\
5       & Rank                       & 1                   & ~Random           & ~Random         & ~Random     & ~Random      \\
6       & Discrete Fourier Transform & 1                   & ~Random           & ~Random         & ~Random     & ~Random      \\
7       & Non-overlapping T. M.      & 1                   & ~Random           & ~Random         & ~Random     & ~Random      \\
8       & Overlapping T.M.           & 1                   & ~Random           & ~Random         & ~Random     & ~Random      \\
9       & Maurer’s Universal         & 1                   & ~Random           & ~Non-Random     & ~Random     & ~Random      \\
10      & Linear complexity          & 1                   & ~Random           & ~Random         & ~Random     & ~Non-Random  \\
11      & Serial                     & 2                   & ~Random           & ~Random         & ~Random     & ~Random      \\
12      & Approximate Entropy        & 1                   & ~Random           & ~Random         & ~Random     & ~Random      \\
13      & Cumulative sums            & 2                   & ~Random           & ~Random         & ~Random     & ~Random      \\
14      & Random Excursions          & 8                   & ~Random           & ~Random         & ~Random     & ~Random      \\
15      & Random Excursions Variant  & 18                  & ~Random           & ~Random         & ~Random     & ~Random     \\ \hline

\end{tabular}
\refstepcounter{table}\label{table:tableS1}
\end{table}

\begin{figure}[t!]
    \centering
    \includegraphics[width=0.90\textwidth]{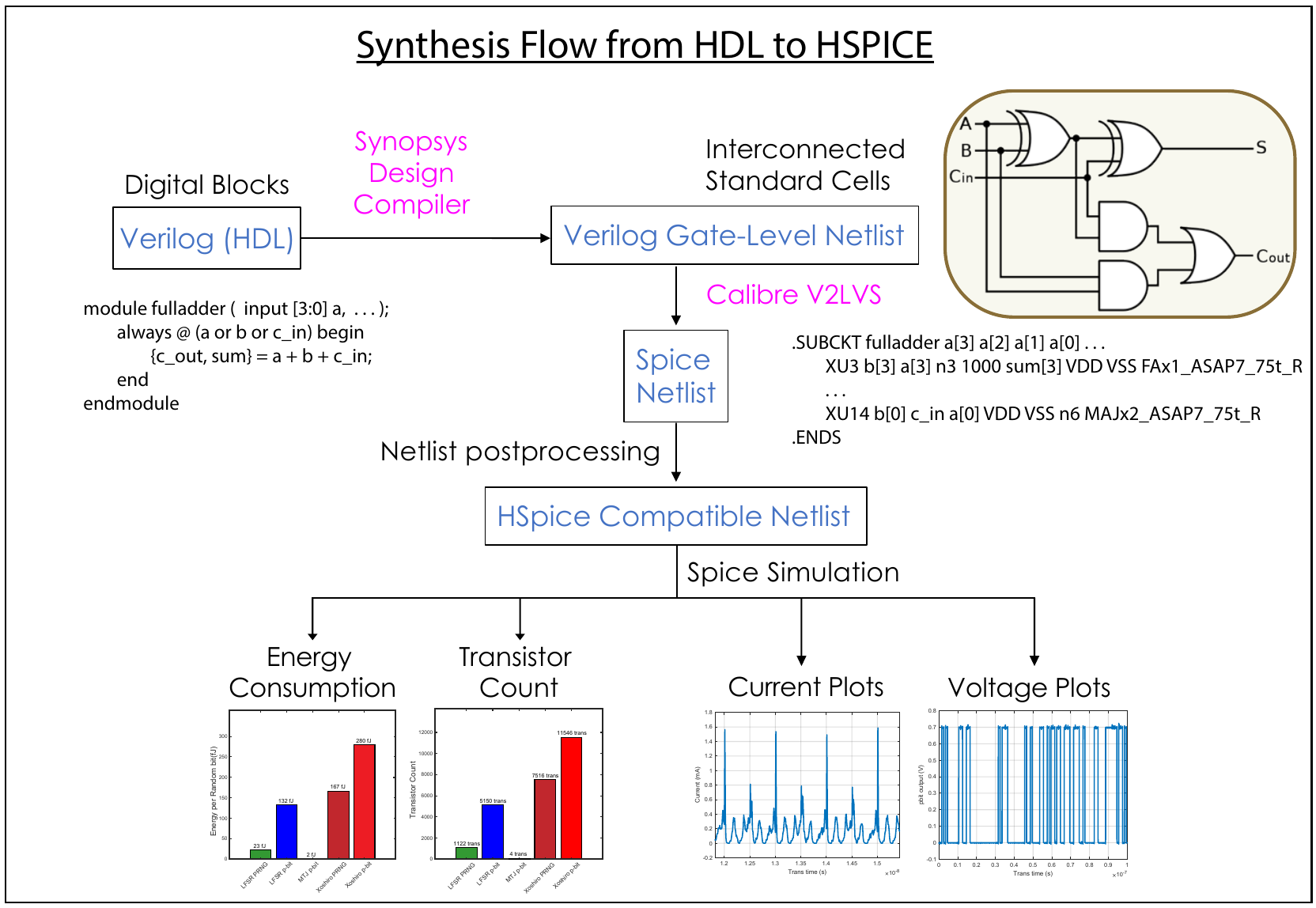}
    % \vspace{10pt} 
    \justify{\textbf{\ref{fig:FigS6}. }Flowchart highlighting the key steps in the digital synthesis flow from hardware description languages such as Verilog all the way down to SPICE.}
     \refstepcounter{figure}\label{fig:FigS6}
     \vspace{-10pt} 
 \end{figure}

\section{Synthesis Flow }
\label{sec:synth}
The synthesis process followed involves the conversion of HDL codes modeling the digital p-bit to a functionally equivalent SPICE netlist (~\ref{fig:FigS6}). This HDL-to-SPICE conversion first involves using Synopsys Design Compiler (DC) to generate the HDL gate-level netlist from the open-source ASAP7 PDK library files. Subsequently, we used Calibre’s Verilog-to-LVS (V2LVS) tool to translate the gate-level netlist into a SPICE compatible netlist. After post-processing using a custom Mathematica script to get the netlist HSPICE compatible, we run transient simulations to obtain energy consumption, transistor counts, and current and voltage plots. With Synopsys DC, we used the regular voltage threshold (RVT) database files, and with V2LVS we used the RVT CDL files. For the HSPICE simulations, a VDD value of 0.7 V was used, and clock frequencies from 100 MHz to 1 GHz were tested, with 1 GHz being used for all results reported in this work.

 \section{Functional verification of p-bits and PRNGs}
\label{sec:funcver}
\begin{figure}[htbp]
    \centering
    \includegraphics[width=0.99\textwidth]{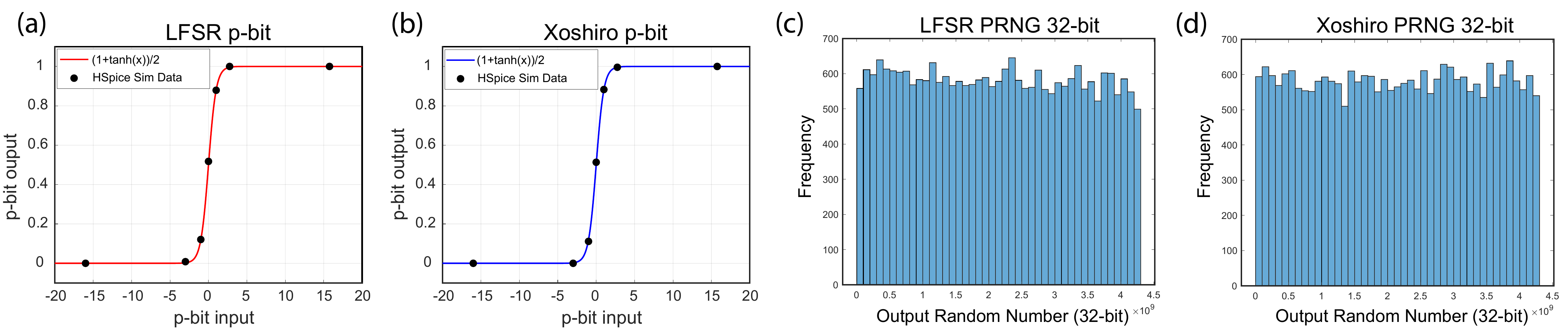}
    % \vspace{10pt} 
    \justify{\textbf{\ref{fig:FigS7}. }(a) Verifying the functionality of 32-bit LFSR and (b) Xoshiro-based p-bits synthesized using ASAP7 PDK. Input current is converted from the 2's complement s43 representation to decimal equivalent. The y-axis indicates the probability of the p-bit being in a 1 state calculated over 2500 clock cycles of a transient p-bit simulation generating random bits. (c) Verifying the functionality of 32-bit LFSR and (d) Xoshiro PRNGs synthesized using ASAP7 PDK. 32-bit binary outputs were mapped to their decimal equivalent, and a 25000 clock cycle transient simulation generated random 32-bit values where $f_{\rm clk} = 1$ GHz.}
     \refstepcounter{figure}\label{fig:FigS7}
 \end{figure}

\begin{figure}[htbp]
    \centering
    \includegraphics[width=0.80\textwidth]{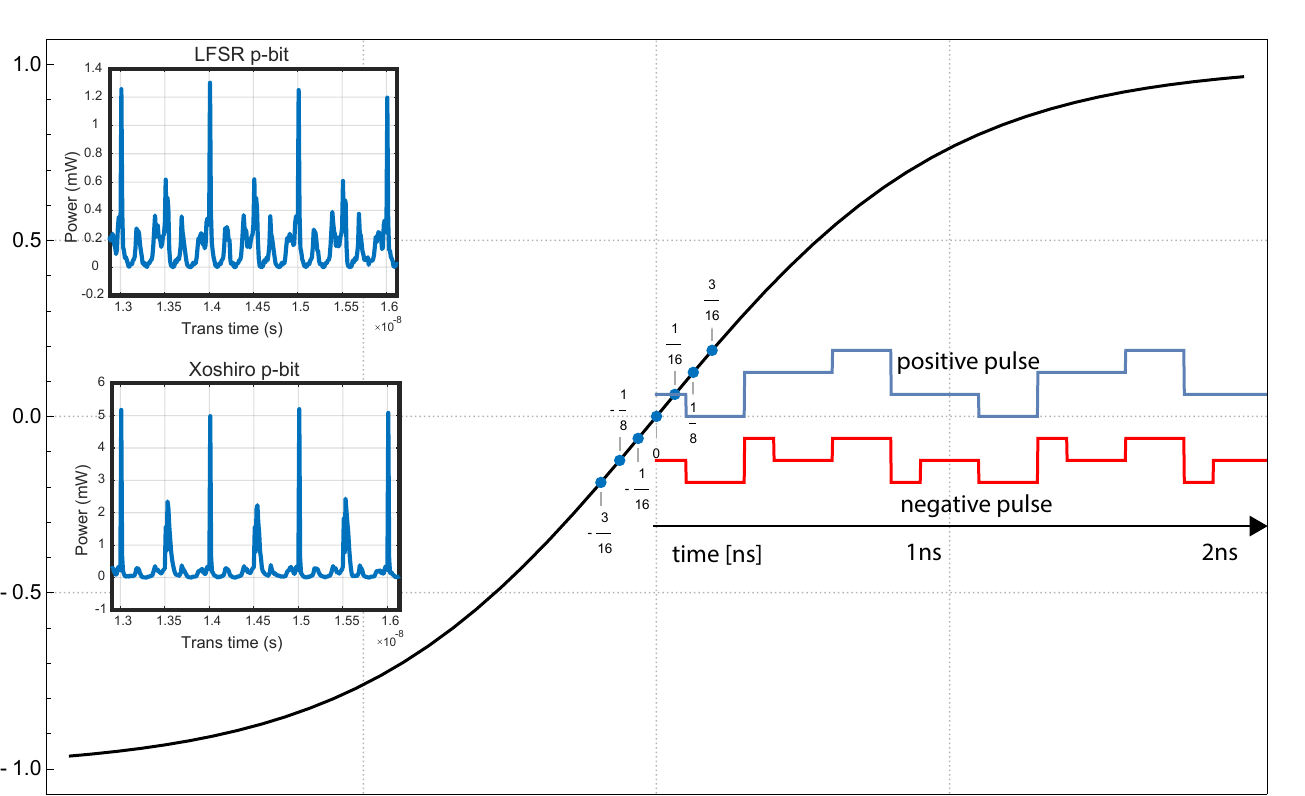}
    % \vspace{10pt} 
   \justify{\textbf{\ref{fig:FigS8}. }Input profile to activate the LUT: 1 GHz positive and negative pulse are applied to the input of the p-bit to activate LUT transistors. Insets show representative transient simulations for instantaneous power consumption for LFSR and Xoshiro-based p-bits.} 
     \refstepcounter{figure}\label{fig:FigS8}
 \end{figure}

As seen from ~\ref{fig:FigS7}a,b, the experimental data obtained from HSPICE by varying the p-bit inputs (8-bit LUT input) of the synthesized circuit falls on the theoretically expected \({1+\mathrm{tanh}(x)}{/2}\)  for both LFSR and Xoshiro-based p-bits. Varying the p-bit input allows us to tune the probability with which the p-bit fluctuates in accordance with the sigmoid seen in ~\ref{fig:FigS7}a,b. In ~\ref{fig:FigS7}c,d, we observe that across 25000 experimental samples, the distribution of outputs obtained from the digitally synthesized PRNG (LFSR or Xoshiro) are uniformly random. These two results verify the functionality of the digital p-bit synthesized using ASAP7 by a transistor-level simulation performed in HSPICE. 

\section{Power analysis of p-computers}
\label{sec:energy} 
\subsection{Synapse power} 
It is important to note that our calculations do not include any power analysis for the synapse (Supplementary Eq.~\eqref{eq:syn}) so far. Earlier estimations \cite{sutton2020autonomous} indicated that the synapse power is at least 50\% of the overall power consumption. Depending on the implementation, for example, using analog crossbars or in-memory computing techniques, the synapse power could show a large degree of variation and we do not explore these possibilities in this paper. 

\subsection{Digital p-bit power}

Insets of ~\ref{fig:FigS8} show high resolution, representative section of the power plots for LFSR and Xoshiro p-bits. Energy analysis was performed by integrating the power over the transient time followed by averaging over 100 clock cycles of a 1 GHz clock, using trapezoidal numerical integration ($\rm trapz$) in MATLAB.

In order to estimate the energy contribution of the LUT to the energy of generating a random bit, we vary the least significant 3 bits of the p-bit input to keep the LUT actively switching to simulate normal operating conditions during probabilistic computations. 
We do this in two different ways by generating 2 switching sequences of the form 00000xxx and 11111xxx, where ``x'' switches between 0 and 1 (see ~\ref{fig:FigS8} for the pulse shapes). The first switching pattern is shown by the positive pulse in blue, where the LUT is traversing the sigmoid just above the zero point. The second switching pattern shown by the negative pulse in red has the LUT switching between values just below the zero point. We choose these input variations to have a 1 GHz frequency and measure the average power and energy dissipation over 100 clock cycles. For the positive pulse, the results are shown in FIG.~\ref{fig:Fig4}b. For the negative pulse we obtain an energy consumption for a 32-bit LFSR p-bit as 119 fJ, and that of a 32-bit Xoshiro p-bit as 267 fJ, similar to what we observed for the positive pulse, reported in the main text, FIG.~\ref{fig:Fig4}b.

\section{Benchmarking and projection of p-bits: Roadmap}
\label{sec:roadmap} 

In this section, we provide projections and benchmarks for probabilistic inference and sampling hardware at device, circuit and systems levels which are summarized in Supplementary Table 4, below. \vspace{3pt}

\noindent \textbf{Device-level:} At the sMTJ level, a key performance metric is the speed of fluctuations which can be measured by autocorrelation and magnetic relaxation times. Here, we seek order-of-magnitude estimates and represent average fluctuations by a single number $\tau$ for simplicity. There are two main methods to design sMTJs, one by employing nanomagnets with perpendicular anistropy (PMA) and another by employing nanomagnets with in-plane anisotropy (IMA). PMA magnets of the type we use in this paper typically have slow fluctuation rates compared to IMA \cite{kaiser2019subnanosecond,kanai2021theory}. PMA magnets whose energy barriers are around 10-15 $k_B T$ fluctuate in the millisecond range \cite{borders2019integer,kaiser2022hardware}. On the other hand, recent device-level experiments using IMA magnets have shown fluctuations with 1-10 nanosecond rates \cite{hayakawa2021nanosecond,safranski2021demonstration,schnitzspan2023nanosecond}. sMTJs made out of IMA magnets possess other favorable properties such as bias-independence \cite{camsari2021double} to build robust p-bit circuits. \vspace{3pt}

\noindent \textbf{Circuit-level:} Despite advances at the individual device level, p-bit circuits using sMTJs have not yet caught up with the fastest sMTJs. Even though we use a different current-mirror topology for the p-bit circuit in this paper, similar to earlier work, our p-bits use PMA MTJs with milisecond fluctuations. A recent report showed the fastest p-bit circuits with microsecond fluctuations, using IMA magnets \cite{singh2023IEDM}. Reaching nanosecond fluctuations with p-bit circuits might require integrated solutions, of the type that are sought initially in Ref.~\cite{daniel2023experimental}. Given that CMOS circuitry could operate at picosecond timescales, there should be no fundamental obstacles to building GHz p-bit circuits.  \vspace{3pt}

\noindent \textbf{System-level:} At the system level, two main performance metrics for p-computers have been identified. One is the number of nodes in the network, $N$. The other one is sampling throughput that measures the number of probabilistic decisions taken by a system. Highly optimized standalone GPUs/TPUs of similar size graphs provide a sampling throughput of 10 flips/ns, establishing an optimized conventional baseline \cite{block2010multi, preis2009gpu, yang2019high, fang2014parallel}. These chips consume around 100W of power. Given this GPU/TPU background, we provide established experimental data and projections for CMOS and CMOS+sMTJ-based probabilistic computers. 

The FPGA columns represent our fully digital FPGA work with different quality RNGs (LFSR and Xoshiro). Depending on the bit precision and network connectivity, these digital solvers can sustain up to $N=10^4$ p-bits in hardware \cite{aadit2022massively,niazi2023training}. Running in parallel with around 10 MHz frequencies, they reach 100 flips/ns \cite{aadit2022massively} in sampling throughput, about an order of magnitude higher than typical GPU and TPUs.  On the other hand, the sMTJs with perpendicular magnetic anisotropy (PMA) used in our present heterogeneous p-computers currently have fluctuation times around $\tau=1$ ms. However, near-term projections with  $N=10^4$ p-bits using sMTJs with in-plane magnetic anisotropy (IMA) ($\tau\approx1$ns  \cite{safranski2021demonstration,hayakawa2021nanosecond,schnitzspan2023nanosecond}) can reach $10^4$ flips/ns in sampling throughput. In the case of heterogeneous p-computers driven by sMTJs, the external p-bits can drive a large number of \textit{digital} p-bits inside the FPGA. In such a case, around N=10,000 digital p-bits can be driven by sMTJs and the sampling throughput would be limited by the synapse time inside the FPGA (or the digital ASIC). This number shows that even in this modest scale, heterogeneous computers can already provide computational advantages over-optimized typical TPU/GPUs. Ultimate, fully-integrated and sMTJ-based computers with $N=10^6$ p-bits can reach sampling throughputs of $10^6$ flips/ns, 5 orders of magnitude faster than typical TPU/GPUs. TPU/GPU references discussed have been plotted on a power consumption versus sampling throughput scale in~\cite{10019530} and \cite{singh2023IEDM}. 

For sampling throughput projections at large scale, we use $N/\tau$, assuming each p-bit flips independently of each other. This basic formula assumes that each flip is communicated to neighboring p-bits \textit{before} a new flip is attempted, as otherwise flips may not be useful. If the network density scales as $O(k\cdot N)$ with some small $k$ corresponding to sparse networks, the fast communication assumption is warranted and ideal parallelism can be achieved. 

For FPGA-based p-computers, the total power consumption is around 30W, including peripheral and unrelated circuits beyond the digital synthesis of our design. For heterogeneous computers of the type we consider in this work (5 MTJs + FPGA), the total power is also dominated by the FPGA power and is also around 30W.  Unlike the present work where we used sMTJs to clock digital PRNGs such as LFSRs, in the future we envision the sMTJ-based analog p-bits as standalone blocks interacting through a CMOS underlayer without any seeding of CMOS PRNGs. In terms of power estimates for such fully sMTJ-based p-computers, detailed circuit simulations with experimentally established parameters indicate a power consumption of 10 $\mu$W per p-bit \cite{hassan2021quant}. For fully sMTJ-based p-computers with $N=10^6$, this would indicate a total p-bit power of 10 W, with an additional 10 W estimated synapse power \cite{sutton2020autonomous, singh2023IEDM}. Considering how present day MRAM technology has been scaled up to 1 Gbit densities \cite{lee20191gbit}, integrating about $10^6$ p-bits on top of CMOS should be reasonable.

\begin{figure}[h!]
    \centering
    \includegraphics[width=0.99\textwidth]{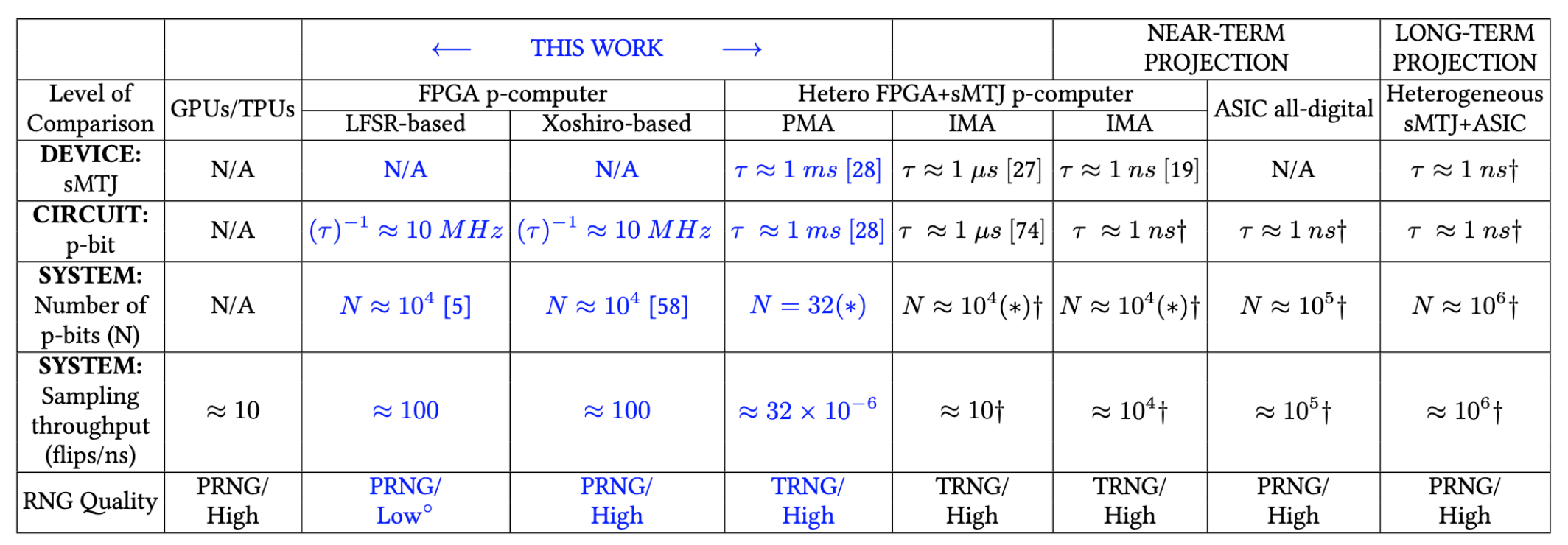}
    \captionsetup{type=table}
       \justify{\textbf{\ref{tab:bench-proj}.} Benchmarking probabilistic hardware from device, circuit and system perspectives.]{Benchmarking probabilistic hardware from device, circuit and system perspectives. For device comparisons, we focus on experimentally demonstrated sMTJ fluctuations. p-bits are circuits that use sMTJs to produce binary stochastic neurons with tunable probability with fluctuations at $\tau^{-1}$ rates. At the system level, we focus on the number of p-bits in a network (N) and sampling throughput, which is given by N/$\tau$, for asynchronous systems with fast synapses computing Supplementary Eq.~\ref{eq:syn} (see text). Also at the system level, we report published data for GPUs/TPUs handling similar probabilistic sampling tasks. Projections are shown using ($\dagger$). (*) In heterogeneous computers of the type we consider in this paper, external sMTJ-based p-bits can drive a large number of digital p-bits in an FPGA or an ASIC. $^{\circ}$ RNG quality is deemed low / high for \textit{sampling} problems rather than combinatorial optimization problems for which LFSR-based PRNGs seem sufficient \cite{aadit2022massively}.}}  
         \refstepcounter{table}  \label{tab:bench-proj}
\end{figure}

\end{document}